\documentclass[11pt]{article}

\usepackage{amsmath,amssymb,amsthm,graphicx,slashed,subfig}%
\usepackage{amsmath}%
\usepackage{amsfonts}%
\usepackage{amssymb}%

\usepackage[utf8]{inputenc}
\usepackage[english]{babel}

\usepackage{hyperref}

\pdfoutput=1

\usepackage{graphicx}
\providecommand{\U}[1]{\protect\rule{.1in}{.1in}}
\newtheorem{thm}{Theorem}

\newtheorem{lma}[thm]{Lemma}

\newtheorem{prop}[thm]{Proposition}
\newtheorem{defn}[thm]{Definition}

\newtheorem{rem}[thm]{Remark}

\def\A{\mathcal{A}}

\def\bar{\overline}

\def\C{\mathbb{C}}

\def\dirac{D}

\def\ev{\mathrm{ev}}
\def\epsilon{\varepsilon}

\def\GUT{\mathrm{GUT}}
\DeclareMathOperator{\GeV}{GeV}

\def\bH{\mathbb{H}}
\def\H{\mathcal{H}}

\def\L{\mathcal{L}}

\def\nn{\nonumber}

\def\R{\mathbb{R}}

\newcommand{\Sub}[1]{_{\scriptscriptstyle#1}}

\DeclareMathOperator{\tr}{Tr}
\def\U{\mathcal{U}}
\def\Z{\mathbb{Z}}

\newcommand{\mattwo}[4]{
  \left(\!\!\!\begin{array}{c@{~}c}#1&#2\\#3&#4\\\end{array}\!\!\!\right)
}
\newcommand{\vectwo}[2]{
  \left(\!\!\!\begin{array}{c}#1\\#2\\\end{array}\!\!\!\right)
}

\title{A survey of spectral models of gravity \\coupled to matter}
\author{Ali H. Chamseddine$^{1,2}$ and Walter D. van Suijlekom$^{2}$\\[4mm]
  \begin{minipage}{\linewidth}
    \centering
  $^{1}$Physics Department, American University of Beirut, Lebanon\\
$^{2}$Institute for Mathematics, Astrophysics and Particle Physics, Radboud University Nijmegen, Heyendaalseweg 135, 6525 AJ Nijmegen, The Netherlands.
\\
\bigskip
\texttt{chams@aub.edu.lb, waltervs@math.ru.nl}
\end{minipage}
}

\begin{document}
\maketitle

\begin{center}
  Dedicated to Alain Connes
  \end{center}

\begin{abstract}
This is a survey of the historical development of the Spectral Standard Model
and beyond, starting with the ground breaking paper of Alain Connes in 1988
where he observed that there is a  link between Higgs fields and finite
noncommutative spaces. We present the important contributions that helped in
the search and identification of the noncommutative space that characterizes the fine
structure of space-time. The  nature and properties of the noncommutative
space are arrived at by independent routes and show the uniqueness of the
Spectral Standard Model at low energies and the Pati--Salam unification model
at high energies. 

  \end{abstract}

\tableofcontents

\section{Introduction}
In 1988, at the height of the string revolution, there appeared an alternative
way to think about the structure of space-time, based on the breathtaking
progress in the new field of noncommutative geometry. Despite the success of
string theory in incorporating gravity, consistency of the theory depended on
the existence of supersymmetry as well as six or seven extra dimensions.
Enormous amount of research was carried out to obtain the Standard Model from
string compactification, which even up to day did not materialize. Most
compactifications start in ten dimensions with the Yang--Mills gauge group
$E_{8}\times E_{8}$ requiring a very large number of fields to become massive
at high energies. In a remarkable paper, Alain Connes laid down the blue
print of a new innovative approach to uncover the origin of the Standard Model
and its symmetries \cite{C90}. The foundation of this approach was based on
noncommutative geometry, a field he founded few years before \cite{C85} (see also \cite{C94}). Alain realized
that by making space slightly noncommutative by tensoring the four dimensional
space with a space of two points, one gets a parallel universe where the
distance between the two sheets is of the order of $10^{-16}$ cm, with the
unexpected bonus of having the Higgs scalar field mediating between them.
Although this looked similar to the idea of Kaluza--Klein, there were essential
differences, mainly in avoiding the huge number of the massive tower of states
as well as obtaining the Higgs field in a representation which is not the
adjoint. Soon after this work inspired similar approaches also based on
extending the four-dimensional space to become noncommutative \cite{Dub88,DKM89,DKM89b,DKM90,CFF92}.

\bigskip

In this survey we will review the key developments that allowed noncommutative geometry to
deepen our understanding of the structure of space-time and explain from first
principles why and how nature dictates the existence of the elementary
particles and their fundamental interactions. In Section 2 we will start by
reviewing the pioneering work of Alain Connes \cite{C90} introducing the
basic mathematical definitions and structures needed to define a
noncommutative space. We summarize the characteristic ingredients in the
construction of the Connes--Lott model and later generalizations by others. We
then consider how to develop the analogue of Riemannian geometry for
non-commutative spaces, and to incorporate the gravitational field in the
Connes--Lott model. In Section 3 we present a breakthrough in the development
of noncommutative geometry with the introduction of the reality operator which
led to the definition of KO dimension of a noncommutative space. With this it
became possible to present the reconstruction theorem of Riemannian geometry
from noncommutative geometry. Section 4 covers the formulation and
applications of the spectral action principle where the spectrum of the Dirac
operator plays a dominant role in the study of noncommutative spaces. This key
development allowed to obtain the dynamics of the Standard Model coupled to
gravity in a non-ambiguous way, and to study geometric invariants of
noncommutative spaces. We then show that incorporating right-handed
neutrinos with the fundamental fermions forces a change in the algebra of the
noncommutative space and the use of real structures to impose simultaneously
the reality and chirality conditions on fundamental states, singling out the
KO dimension to be 6. We show in detail how the few requirements
about KO dimension, Majorana masses for right-handed neutrinos and the first
order condition on the Dirac operator, singles out the geometry of
the Standard Model. In Section 5 we present a classification of finite
noncommutative spaces of KO dimension 6 showing the almost uniqueness of the
Standard spectral model. In Section 6 we give a prescription of constructing
spectral models from first principles and show that the spectral Standard
Model agrees with the available experimental limits, provided that the scale
giving mass to the right-handed neutrinos is promoted to a singlet scalar
field. We then show that there exists a more general case where the
first order condition on the Dirac operator is removed, the singlet scalar
fields become part of a larger representation of the Pati--Salam model. The
Standard Model becomes a special point in the spontaneous breaking of the
Pati--Salam symmetries. In Section 6 we show that a different starting point
where a Heisenberg like quantization condition in terms of the Dirac operator
considered as momenta and two possible Clifford-algebra valued maps from the
four-dimensional manifold to two four-spheres $S^{4}$ result in noncommutative
spaces with quantized volumes. The Pati--Salam model and its various truncations are
uniquely determined as the symmetries of the spaces solving the constraint.
Section 7 contains the conclusions and a discussion of possible directions of future research.

\subsubsection*{Acknowledgements}
The work of A. H. C. is supported in part by the National Science Foundation
Grant No. Phys-1518371. He also thanks the Radboud Excellence Initiative for
hosting him at Radboud University where this research was carried out. We would
like to thank Alain Connes for sharing with us his insights and ideas.

\section{Early days of the spectral Standard Model}
The first serious attempt to utilize the ideas of noncommutative geometry in
particle physic was made by Alain Connes in 1988 in his paper "Essay on
physics and noncommutative geometry" \cite{C90}. He observed that it is possible to
change the structure of the (Euclidean) space-time so that the action
functional gives the Weinberg-Salam model. The main emphasis was on the
conceptual understanding of the Higgs field, which he calls, the black box of
the standard model. The qualitative picture was taken to be of a two-sheeted
Euclidean space-time separated by a distance of the order of $10^{-16}$ cm. In
order to simplify the presentation, and to easily follow the historical
development, we will use a uniform notation, representing old results in a new
format. It is therefore more efficient to start with the basic definitions.

\subsection{Noncommutative spaces and differential calculus}
A noncommutative space is determined from the spectral data $\left(
\mathcal{A},\mathcal{H},D,\gamma,J\right)  $ where $\mathcal{A}$ is an
associative algebra with unit element $1$ and involution *, $\mathcal{H}$ a
Hilbert space carrying a faithful representation $\pi$ of the algebra, $D$ is
a self-adjoint operator on $\mathcal{H}$ with $\left(  D^{2}+1\right)  ^{-1}$
compact, $\gamma$ is the unitary chirality operator and $J$ an anti-unitary
operator on $\mathcal{H}$, the reality structure. The operator $J$ was
introduced later in 1994 \cite{C95}.

In the model proposed in 1988, there were ambiguities in defining the algebra
and the action on the Hilbert space. These were rectified in the 1990 paper \cite{CL91}
with John Lott, in what became known as the Connes--Lott model. In order to
appreciate the enormous progress made over the years, we will summarize this
model in a simplified presentation. A more detailed account can be found in the early reviews \cite{VG93,MGV98,Kas93,Kas96,KasS96,KasS97}. Note that at around the same time a derivation based differential calculus was introduced by others in \cite{Dub88,DKM89,DKM89b,DKM90} with many similarities to the model proposed by Connes in 1988. 

We first need to first
introduce new ingredients. Given a unital involutive algebra $\mathcal{A}$,
the universal differential algebra over $\mathcal{A}$ is defined as
\begin{equation*}
\Omega^{\ast}\left(  \mathcal{A}\right)  =%
{\displaystyle\bigoplus\limits_{n=0}^{\infty}}
\Omega^{n}\left(  \mathcal{A}\right)
\end{equation*}
where we set $\Omega^{0}\left(  \mathcal{A}\right)  =\mathcal{A}$, and take%
\begin{equation*}
\Omega^{n}\left(  \mathcal{A}\right)  =\left\{
{\displaystyle\sum\limits_{i}}
a_{0}^{i}da_{1}^{i}da_{2}^{i}\cdots da_{n}^{i},\qquad a_{j}^{i}\in
\mathcal{A},\quad\forall i,j\right\}  ,\quad n=1,2,\cdots
\end{equation*}
Here $da$ denotes an equivalence class of $\mathcal{A}$, modulo the following
relations
\begin{equation*}
d\left(  a\cdot b\right)  =da\cdot b+a\cdot db,\qquad d1=0,\qquad d^{2}=0
\end{equation*}
An element of $\Omega^{n}\left(  \mathcal{A}\right)  $ is called a form of
degree $n.$ One forms can be considered as connections on a line bundle whose
space of sections is given by the algebra $\mathcal{A}$. A one form $\rho
\in\Omega^{1}\left(  \mathcal{A}\right)  $ is expressed in the form%
\begin{equation*}
\rho=%
{\displaystyle\sum\limits_{i}}
a^{i}db^{i},\qquad a^{i},b^{i}\in\mathcal{A}%
\end{equation*}
and since $d1=0,$ we may impose the condition $%
{\displaystyle\sum\limits_{i}}
a^{i}b^{i}=1,$ without any loss in generality. We say that $\left(
\mathcal{H},D\right)  $ is a Dirac K-cycle for $\mathcal{A}$ if and only if there exists
an involutive representation $\pi$ of $\mathcal{A}$ on $\mathcal{H}$
satisfying $\pi\left(  a\right)  ^{\ast}=\pi\left(  a^{\ast}\right)  $ with
the properties that $\pi\left(  a\right)  $ and $\left[  D,\pi\left(
a\right)  \right]  $ are bounded operators on $\mathcal{H}$ for all
$a\in\mathcal{A}$. The K-cycle is called even if there exists a chirality
operator $\gamma$ such that $\gamma D=-D\gamma,$ $\gamma=\gamma^{-1}%
=\gamma^{\ast}$ and $\left[  \gamma,\pi\left(  a\right)  \right]  =0,$
otherwise it is odd. The action of $\pi$ on $\Omega^{\ast}\left(
\mathcal{A}\right)  $ is defined as
\begin{equation*}
\pi\left(
{\displaystyle\sum\limits_{i}}
a_{0}^{i}da_{1}^{i}\cdots da_{n}^{i}\right)  =%
{\displaystyle\sum\limits_{i}}
\pi\left(  a_{0}^{i}\right)  [D,\pi\left(  a_{1}^{i}\right)  ]\cdots\lbrack
D,\pi\left(  a_{n}^{i}\right)  ]
\end{equation*}
The space of auxiliary fields is defined by
\begin{equation*}
\mathrm{Aux}=\ker\pi+d\,\ker\pi
\end{equation*}
where
\begin{equation*}
\ker\pi=%
{\displaystyle\bigoplus\limits_{n=0}^{\infty}}
\left\{
{\displaystyle\sum\limits_{i}}
a_{0}^{i}da_{1}^{i}\cdots da_{n}^{i}\,:\pi\left(
{\displaystyle\sum\limits_{i}}
a_{0}^{i}da_{1}^{i}\cdots da_{n}^{i}\right)  =0\right\}
\end{equation*}
and
\begin{equation*}
d\ker\pi=%
{\displaystyle\bigoplus\limits_{n=0}^{\infty}}
\left\{
{\displaystyle\sum\limits_{i}}
da_{0}^{i}da_{1}^{i}\cdots da_{n}^{i}\,:\pi\left(
{\displaystyle\sum\limits_{i}}
a_{0}^{i}da_{1}^{i}\cdots da_{n}^{i}\right)  =0\right\}
\end{equation*}
The integral of a form $\alpha\in\Omega^{\ast}\left(  \mathcal{A}\right)  $
over a noncommutative space of metric dimension $d$ is defined \ by setting%
\begin{equation*}%
{\displaystyle\int}
\alpha=\mathrm{Tr}_{w}\left(  \pi\left(  \alpha\right)  D^{-d}\right)
\end{equation*}
where $\mathrm{Tr}_{w}$ is the Dixmier trace.

\subsection{Two-sheeted spacetime}
A simple extension of space-time is taken as a product of continuous
four-dimensional manifold times a discrete set of two points. The algebra is
$\mathcal{A}=\mathcal{A}_{1}\otimes\mathcal{A}_{2}$ acting on the Hilbert
space $\mathcal{H}=\mathcal{H}_{1}\otimes\mathcal{H}_{2}$ where $\mathcal{A}%
_{1}=C^{\infty}\left(  M\right)  $ and $\mathcal{A}_{2}=M_{2}\left(
\mathbb{C}\right)  \oplus M_{1}\left(  \mathbb{C}\right)  ,$ the algebra of
$2\times2$ and $1\times1$ matrices. The Hilbert space is that of spinors of
the form
\begin{equation*}
L=\left(
\begin{array}
[c]{c}%
l\\
e
\end{array}
\right)
\end{equation*}
where $l$ is a doublet and $e$ is a singlet. The spinor $L$ satisfies the
chirality condition $\gamma_{5}\otimes\Gamma_{1}L=L$ where $\Gamma
_{1}=\mathrm{diag}\left(  1_{2},-1\right)  $ is a grading operator. From this
we deduce that $l$ is a left-handed spinor and $e$ is right handed, and we
thus write $l=\left(
\begin{array}
[c]{c}%
\nu_{L}\\
e_{L}%
\end{array}
\right)  $ and $e=e_{R}.$ The Dirac operator is given by $D=D_{1}%
\otimes1+\gamma_{5}\otimes D_{2}$ where $D_{1}=\gamma^{\mu}\partial_{\mu}$ and
$D_{2}$ is the Dirac operator on $\mathcal{A}_{2}$ such that
\begin{equation*}
D_{l}=\left(
\begin{array}
[c]{cc}%
\gamma^{\mu}\partial_{\mu}\otimes1_{2}\otimes1_{3} & \gamma_{5}\otimes
M_{12}\otimes k\\
\gamma_{5}\otimes M_{21}\otimes k^{\ast} & \gamma^{\mu}\partial_{\mu}%
\otimes1\otimes1_{3}%
\end{array}
\right)
\end{equation*}
where $M_{21}=M_{12}^{\ast}$ and $k$ is a $3\times3$ family mixing matrix
representing Yukawa couplings for the leptons. The $1\times2$ matrix $M_{12}$
turns out to be the vev of the Higgs field and is taken as $M_{12}=\mu\left(
\begin{array}
[c]{c}%
0\\
1
\end{array}
\right)  =H_{0}.$ The elements $a\in\mathcal{A}$ have the representation
$a=\left(
\begin{array}
[c]{cc}%
a_{1} & 0\\
0 & a_{2}%
\end{array}
\right)  $ where $a_{1},$ $a_{2}$ are  $2\times2$  and $1\times1$ unitary
valued functions. A quick calculation shows that the self-adjoint one-form
$\rho$ has the representation%
\begin{equation*}
\pi_{1}\left(  \rho\right)  =\left(
\begin{array}
[c]{cc}%
A_{1}\otimes1_{3} & \gamma_{5}\otimes H\otimes k\\
\gamma_{5}\otimes H^{\ast}\otimes k^{\ast} & A_{2}\otimes1_{3}%
\end{array}
\right)
\end{equation*}
where
\begin{align*}
A_{1}  & =\gamma^{\mu}%
{\displaystyle\sum\limits_{i}}
a_{1}^{i}\partial_{\mu}b_{1}^{i},\qquad A_{2}=\gamma^{\mu}%
{\displaystyle\sum\limits_{i}}
a_{2}^{i}\partial_{\mu}b_{2}^{i},\\
H  & =H_{0}+%
{\displaystyle\sum\limits_{i}}
a_{1}^{i}H_{0}b_{2}^{i}.
\end{align*}
The quarks are introduced by taking for the finite space a bimodule structure
relating two algebras $\mathcal{A}$ and $\mathcal{B}$ where the algebra
$\mathcal{B}$ is taken to be $M_{1}\left(  \mathbb{C}\right)  \oplus
M_{3}\left(  \mathbb{C}\right)  $ commuting with the action of $\mathcal{A}.$
In addition, the mass matrices in the Dirac operator are taken to be zero when
acting on elements of $\mathcal{B}.$ The one-form $\eta\in\Omega^{1}\left(
\mathcal{B}\right)  $ has the simple form $B_{1}\mathrm{diag}\left(
1_{2},1\right)  $ where $B_{1}$ is a gauge field associated with $M_{1}\left(
\mathbb{C}\right)  .$ The Hilbert space for the quarks is
\begin{equation*}
Q=\left(
\begin{array}
[c]{c}%
q_{L}\\
u_{R}\\
d_{R}%
\end{array}
\right)  ,\qquad q_{L}=\left(
\begin{array}
[c]{c}%
u_{L}\\
d_{L}%
\end{array}
\right)
\end{equation*}
The representation of $a\in\mathcal{A}$ is $a\rightarrow\left(  a_{1}%
,a_{2},\overline{a}_{2}\right)  $ where $a_{1}$ and $a_{2}$ are a $2\times2$
and $1\times1$ complex valued functions. The Dirac operator acting on the
quark Hilbert space is
\begin{equation*}
D_{q}=\left(
\begin{array}
[c]{ccc}%
\gamma^{\mu}\left(  \partial_{\mu}+\cdots\right)  \otimes1_{2}\otimes1_{3} &
\gamma_{5}\otimes M_{12}\otimes k^{\prime} & \gamma_{5}\otimes\widetilde{M}%
_{12}\otimes k^{^{\prime\prime}}\\
\gamma_{5}\otimes M_{12}^{\ast}\otimes k^{\prime\ast} & \gamma^{\mu}\left(
\partial_{\mu}+\cdots\right)  \otimes1_{3} & 0\\
\gamma_{5}\otimes\widetilde{M}_{12}^{\ast}\otimes k^{^{\prime\prime}\ast} &
0 & \gamma^{\mu}\left(  \partial_{\mu}+\cdots\right)  \otimes1_{3}%
\end{array}
\right)
\end{equation*}
where $k^{\prime}$ and $k^{\prime\prime}$ are $3\times3$ family mixing
matrices and $\widetilde{M}_{12}=\mu\left(
\begin{array}
[c]{c}%
1\\
0
\end{array}
\right)  .$ The one form in $\Omega^{1}\left(  \mathcal{A}\right)  $ has then
the representation
\begin{equation*}
\pi_{q}\left(  \rho\right)  =\left(
\begin{array}
[c]{ccc}%
A_{1}\otimes1_{3} & \gamma_{5}\otimes H\otimes k^{\prime} & \gamma_{5}%
\otimes\widetilde{H}\otimes k^{^{\prime\prime}}\\
\gamma_{5}\otimes H^{\ast}\otimes k^{\prime\ast} & A_{2}\otimes1_{3} & 0\\
\gamma_{5}\otimes\widetilde{H}^{\ast}\otimes k^{^{\prime\prime}\ast} & 0 &
\overline{A}_{2}\otimes1_{3}%
\end{array}
\right)
\end{equation*}
where $\widetilde{H}_{a}=\epsilon_{ab}H^{b}.$ When acting on the algebra
$\mathcal{B}$ the Dirac operator has zero mass matrices and the one-form
$\eta$ in $\Omega^{1}\left(  \mathcal{B}\right)  $ has the representation
$\pi_{q}\left(  \eta\right)  =B_{2}\mathrm{diag}\left(  1_{2},1\right)  $
where $B_{2}$ is the gauge field associated with $M_{3}\left(  \mathbb{C}%
\right)  .$ Imposing the unimodularity condition on the algebras $\mathcal{A}$
and $\mathcal{B}$ would then relate the $U\left(  1\right)  $ factors in both
algebras so that $\mathrm{tr}\left(  A_{1}\right)  =0,$ $A_{2}=B_{1}%
=-\mathrm{tr}\left(  B_{2}\right)  \equiv\frac{i}{2}g_{1}B$. With these we can
then write
\begin{align*}
A_{1}  & =-\frac{i}{2}g_{2}\sigma^{a}A_{a}\\
B_{2}  & =-\frac{i}{6}g_{1}B-\frac{i}{2}g_{3}V^{i}\lambda_{i}%
\end{align*}
where $g_{1},$ $g_{2}$ and $g_{3}$ are the $U\left(  1\right)  ,$ $SU\left(
2\right)  $ and $SU\left(  3\right)  $ gauge coupling constants,  $\sigma^{a}$
and $\lambda^{i}$ are the Pauli and Gell-Mann matrices respectively. The
fermionic actions for the leptons and quarks are then given by
\begin{align*}
\left\langle L,\left(  D+\rho+\eta\right)  L\right\rangle  & =%
{\displaystyle\int}
d^{4}x\sqrt{g}\left(  \overline{L}\left(  D_{l}+\pi_{l}\left(  \rho\right)
+\pi_{l}\left(  \eta\right)  \right)  L\right)  \\
\left\langle Q,\left(  D+\rho+\eta\right)  Q\right\rangle  & =%
{\displaystyle\int}
d^{4}x\sqrt{g}\left(  \overline{Q}\left(  D_{q}+\pi_{q}\left(  \rho\right)
+\pi_{q}\left(  \eta\right)  \right)  Q\right)
\end{align*}
These terms can be easily checked to reproduced all the fermionic terms of the
Standard Model. 

The bosonic action is the sum of the square of curvatures in both the lepton
and quark sectors. These are given by
\begin{align*}
I_{l}  & =\mathrm{Tr}\left(  C_{l}\left(  \theta_{\rho}+\theta_{\eta}\right)
^{2}D_{l}^{-4}\right)  \\
I_{q}  & =\mathrm{Tr}\left(  C_{q}\left(  \theta_{\rho}+\theta_{\eta}\right)
^{2}D_{q}^{-4}\right)
\end{align*}
where
\begin{equation*}
\theta_{\rho}\equiv d\rho+\rho^{2}%
\end{equation*}
is the curvature of $\rho,$ and $C_{l}$ and $C_{q}$ are constant elements of
the algebra. Since the representation $\pi$ has a kernel, the auxiliary fields
must be projected out. This step mainly affects the potential. After some
algebra one can show that the bosonic action given above reproduces all the
bosonic interactions of the Standard Model with the same number of parameters.
If one assumes that  $C_{l}$ and $C_{q}$ belong to the center of the algebra,
then one can get fixed values for the top quark mass and Higgs mass. The main
advantage of the noncommutative construction of the Standard Model is that one
gets a geometrical understanding of the origin of the Higgs field and a
unification of the gauge and Higgs sectors. One sees that the Higgs fields are
the components of the one form along discrete directions.

\subsection{Constructions beyond the Standard Model}
The early constructions of the Standard Model provided encouragements to look
further into noncommutative spaces. The construction was also complicated with
some ambiguities such as the independence of the lepton and quark sectors, the
construction of the Higgs potential and projecting out the auxiliary fields.
It was then natural to ask whether it is possible to go beyond the Standard
Model. In particle physics the route taken was to consider larger groups such
as $SU\left(  5\right)  $ or $SO(10)$ which contains  $U\left(  1\right)
\times SU\left(  2\right)  \times SU\left(  3\right)  $ as a subgroup. The
main advantage of GUT is that the fermionic fields are unified in one or two
representations, the most attractive possibility being $SO(10)$ where the
spinor representation $16_{s}$ contains all the known fermions in addition to
the right-handed neutrino. The simplicity in the fermionic sector did not make
the theory more predictive because of the arbitrariness of the Higgs sector.
There are many possible Higgs representations that can break the symmetry
spontaneously from $SO(10)$ to $SU\left(  3\right)  \times U\left(  1\right)
.$ In the noncommutative construction the Higgs sector is more constrained
which was taken as an encouragement to explore the possibility of considering
larger matrix algebras. As an example if one arranges the leptons in the form
$L=\left(
\begin{array}
[c]{c}%
l_{L}\\
l_{R}%
\end{array}
\right)  $ where $l=\left(
\begin{array}
[c]{c}%
\nu\\
e
\end{array}
\right)  $ then the corresponding algebra will be $M_{2}\left(  \mathbb{C}%
\right)  \oplus M_{2}\left(  \mathbb{C}\right)  .$ A natural possibility is
then to consider a discrete space of four points and where the fermions are
arranged in the format $\psi=\left(
\begin{array}
[c]{c}%
l_{L}\\
l_{R}\\
l_{L}^{c}\\
l_{R}^{c}%
\end{array}
\right)  $ and the representation $\pi$ acting on $\mathcal{A}$ is given by
$\pi\left(  a\right)  =\mathrm{diag}\left(  a_{1},a_{2},\overline{a}%
_{1},\overline{a}_{2}\right)  $ where $a_{1},$ $a_{2}$ are $2\times2$ complex
matrices. The resulting model has $SU\left(  2\right)  _{L}\times SU\left(
2\right)  _{R}\times U\left(  1\right)  _{B-L}$ with the Higgs fields in the
representations $\left(  2,2\right)  ,$ $\left(  3,1\right)  +\left(
1,3\right)  $ of $SU\left(  2\right)  _{L}\times SU\left(  2\right)  _{R}.$ We
can summarize the steps needed to construct noncommutative particle physics
models. First we specify the fermion representations then we choose the number
of discrete points and the symmetry between them. From this we deduce the
appropriate algebra and the map $\pi$ acting on the Hilbert space of spinors.
Finally we write down the Dirac operator acting on elements of the algebra and
choose the mass matrices to correspond to the desired vacuum of the Higgs
fields. 

To illustrate these steps consider the chiral space-time spinors $P_{+}\psi$
to be in the $16_{s}$ representation of $SO(10),$ where $P_{+}$ is the
$SO(10)$ chirality operator, and the number of discrete points to be four. The
Hilbert space is taken to be $\Psi=\left(
\begin{array}
[c]{c}%
P_{+}\psi\\
P_{+}\psi\\
P_{-}\psi^{c}\\
P_{-}\psi^{c}%
\end{array}
\right)  $ where $\psi^{c}=BC\overline{\psi}^{T},$ $C$ being the charge
conjugation matrix while $B$ is the $SO\left(  10\right)  $ conjugation
matrix. The finite algebra is taken to be $\mathcal{A}_{2}=P_{+}\left(
\mathsf{Cliff\,SO}\left(  10\right)  \right)  P_{+},$ and the finite Hilbert
space $\mathcal{H}_{2}=\mathbb{C}^{32}.$ Let $\pi_{0}$ denote the
representation of the algebra $\mathcal{A}$ on the Hilbert space $\mathcal{H}$ and
let $\overline{\pi}_{0}$ denote the anti representation defined by
$\overline{\pi}_{0}\left(  a\right)  =B\overline{\pi_{0}\left(  a\right)
}B^{-1}.$ We then define $\pi\left(  a\right)  =\pi_{0}\left(  a\right)
\oplus\pi_{0}\left(  a\right)  \oplus\overline{\pi}_{0}\left(  a\right)
\oplus\overline{\pi}_{0}\left(  a\right)  .$ The Dirac operator is taken to
be
\begin{equation*}
\left(
\begin{array}
[c]{cccc}%
\gamma^{\mu}\partial_{\mu}\otimes1_{32}\otimes1_{3} & \gamma_{5}\otimes
M_{12}\otimes K_{12} & \gamma_{5}\otimes M_{13}\otimes K_{13} & \gamma
_{5}\otimes M_{14}\otimes K_{14}\\
\gamma_{5}\otimes M_{12}^{\ast}\otimes K_{12}^{\ast} & \gamma^{\mu}%
\partial_{\mu}\otimes1_{32}\otimes1_{3} & \gamma_{5}\otimes M_{23}\otimes
K_{23} & \gamma_{5}\otimes M_{24}\otimes K_{24}\\
\gamma_{5}\otimes M_{13}^{\ast}\otimes K_{13}^{\ast} & \gamma_{5}\otimes
M_{23}^{\ast}\otimes K_{23}^{\ast} & \gamma^{\mu}\partial_{\mu}\otimes
1_{32}\otimes1_{3} & \gamma_{5}\otimes M_{34}\otimes K_{34}\\
\gamma_{5}\otimes M_{14}^{\ast}\otimes K_{14}^{\ast} & \gamma_{5}\otimes
M_{24}^{\ast}\otimes K_{24}^{\ast} & \gamma_{5}\otimes M_{34}^{\ast}\otimes
K_{34}^{\ast} & \gamma^{\mu}\partial_{\mu}\otimes1_{32}\otimes1_{3}%
\end{array}
\right)
\end{equation*}
where the $K_{mn}$ are $3\times3$ family mixing matrices commuting with
$\pi\left(  a\right)  .$ We may impose the exchange symmetries
$1\leftrightarrow2$ and $3\leftrightarrow4$ so that $M_{12}=M_{12}^{\ast
}=\mathcal{M}_{0},$ $M_{13}=M_{14}=M_{23}=M_{24}=\mathcal{N}_{0},$
$M_{34}=M_{34}^{\ast}=B\overline{\mathcal{M}}_{0}B^{-1}.$ Computing
$\pi\left(  \rho\right)  $ we get%
\begin{equation*}
\pi\left(  \rho\right)  =\left(
\begin{array}
[c]{cccc}%
A & \gamma_{5}\mathcal{M}K_{12} & \gamma_{5}\mathcal{N}K_{13} & \gamma
_{5}\mathcal{N}K_{14}\\
\gamma_{5}\mathcal{M}K_{12}^{\ast} & A & \gamma_{5}\mathcal{N}K_{23} &
\gamma_{5}\mathcal{N}K_{24}\\
\gamma_{5}\mathcal{N}^{\ast}K_{13}^{\ast} & \gamma_{5}\mathcal{N}^{\ast}%
K_{23}^{\ast} & B\overline{A}B^{-1} & \gamma_{5}B\overline{\mathcal{M}}%
B^{-1}K_{34}\\
\gamma_{5}\mathcal{N}^{\ast}K_{14}^{\ast} & \gamma_{5}\mathcal{N}^{\ast}%
K_{24}^{\ast} & \gamma_{5}B\overline{\mathcal{M}}B^{-1}K_{34}^{\ast} &
B\overline{A}B^{-1}%
\end{array}
\right)
\end{equation*}
where
\begin{align*}
A  & =P_{+}%
{\displaystyle\sum\limits_{i}}
a^{i}\gamma^{\mu}\partial_{\mu}b^{i}P_{+}\\
\mathcal{M}+\mathcal{M}_{0}  & =P_{+}%
{\displaystyle\sum\limits_{i}}
a^{i}\mathcal{M}_{0}b^{i}P_{+}\\
\mathcal{N}+\mathcal{N}_{0}  & =P_{+}%
{\displaystyle\sum\limits_{i}}
a^{i}\mathcal{N}_{0}B\overline{b}^{i}B^{-1}P_{-}%
\end{align*}
One sees immediately that the Higgs fields $\mathcal{M}$ and $\mathcal{N}$ are
in the $16_{s}\times16_{s}$ and $16_{s}\times\overline{16}_{s}$
representations. Equating the action of $A$ on $\psi$ and $\psi^{c}$ will
reduce it to an $SO\left(  10\right)  $ gauge field. Specifying $\mathcal{M}%
_{0}$ and $\mathcal{N}_{0}$ determines the breaking pattern of $SO\left(
10\right)  .$ One can then proceed to construct the bosonic sector and project
out the auxiliary fields to determine the potential. There are very limited
number of models one can construct. These models, however, will suffer the
same problems encountered in the GUT construction, mainly that of low
unification scale of $10^{14}$ Gev implying fast rate of proton decay which is
ruled out experimentally.

\subsection{Coupling matter to gravity}
The dynamics of the gravitational force is based on Riemannian geometry. It is
therefore natural to study the nature of the gravitational field in
noncommutative geometry. The original attempt \cite{CFF93,CFG95} was based on generalizing the
basic notions of Riemannian geometry, notably the theory of linear connections on differential forms. (Note that an alternative route that takes vector fields as a starting point ends with a derivation based differential calculus as in \cite{Dub88} ({\em cf.} \cite{Mad95}). In line with the Connes--Lott model, we will instead take differential forms as our starting point. For more details we also refer to the exposition in \cite[Sect. 10.3]{Lnd97}). 

First one defines the metric as an inner product on a cotangent space. Then one shows that every cycle over
$\mathcal{A}$ yields a notion of cotangent bundle associated with
$\mathcal{A}$ and a Riemannian metric on the cotangent bundle $\Omega_{D}%
^{1}\left(  \mathcal{A}\right)  .$ With the connection $\nabla$ the Riemann
curvature of $\nabla$ on $\Omega_{D}^{1}\left(  \mathcal{A}\right)  $ is
defined by $R\left(  \nabla\right)  :=-\nabla^{2}$ and the torsion by
$T=d-m\circ\nabla$ where $m$ is the tensor product. Requiring $\nabla$ to be
unitary and the torsion to vanish we obtain the Levi--Civita connection. If
$\Omega_{D}^{1}\left(  \mathcal{A}\right)  $ is a finitely generated module,
then it admits a basis $e^{A},$ $A=1,2,\cdots,N,$ and the connection
$\omega_{B}^{A}\in\Omega_{D}^{1}\left(  \mathcal{A}\right)  $ is defined by
$\nabla e^{A}=-\omega_{B}^{A}\otimes e^{B}.$ The components of the torsion
$T\left(  \nabla\right)  $ are defined by $T^{A}=T\left(  \nabla\right)
e^{A}$ then $T^{A}\in\Omega_{D}^{2}\left(  \mathcal{A}\right)  $ is given by
\begin{equation*}
T^{A}=de^{A}+\omega_{B}^{A}e^{B}%
\end{equation*}
Similarly, components of the curvature $R_{B}^{A}\in\Omega_{D}^{2}\left(
\mathcal{A}\right)  $ satisfy the defining property that $R\left(  \nabla\right)  e^{A}=R_{B}%
^{A}\otimes e^{B}$ so that
\begin{equation*}
R_{B}^{A}=d\omega_{B}^{A}+\omega_{C}^{A}\omega_{B}^{C}.
\end{equation*}
The analogue of the Einstein--Hilbert action is then
\begin{equation*}
I\left(  \nabla\right)  :=\kappa^{-2}\left\langle R_{B}^{A}e^{B}%
,e_{A}\right\rangle
\end{equation*}
where $\kappa^{-1}$ is the Planck scale. Computing this action for the product
space $M_{4}\times Z_{2}$ one finds that
\begin{equation*}
I\left(  \nabla\right)  =2%
{\displaystyle\int\limits_{M}}
d^{4}x\sqrt{g}\left(  \kappa^{-2}r-2\partial_{\mu}\sigma\partial^{\mu}%
\sigma\right)
\end{equation*}
where $r$ is the scalar curvature of the Levi--Civita connection of the
Riemannian manifold $M_{4}$ coupled to a scalar field $\sigma.$ Applying this
construction to the Connes--Lott model is rather involved because the two
sheets are not treated symmetrically, being associated with two different
algebras. The complication arise because the projective module is not free and
the basis $e^{A}$ is constrained. The Einstein--Hilbert action in this case is
given by
\begin{equation*}
I\left(  \nabla\right)  =2%
{\displaystyle\int\limits_{M}}
d^{4}x\sqrt{g}\left(  \kappa^{-2}\frac{3}{2}r-2\left(  3+\lambda\right)
\partial_{\mu}\sigma\partial^{\mu}\sigma+c\left(  \lambda\right)  e^{-2\sigma
}\right)
\end{equation*}
where $\lambda=\mathrm{Tr}\left(  kk^{\ast}\right)  ^{2}-1.$ To understand the
significance of the field $\sigma$, we note that by examining the Dirac
operator one finds that the field $\phi=e^{-\kappa\sigma}$ now replaces the
weak scale. Thus quantum corrections to the classical potential will depend on
$\sigma,$ thus the vev of $\sigma$ could be determined from the minimization equations.

\section{The spectral action principle}

Despite the success of the Connes--Lott model and the generalizations that
followed in giving a geometrical meaning to the Higgs field and unifying it
with the gauge fields, it was felt that the construction is not satisfactory.
The first unpleasant feature was the use of the bimodule structure to
introduce the  $SU\left(  3\right)  $ symmetry and the second is the use of
unimodularity condition to get the correct hypercharge assignments to the
particles. Another major problem was the existence of mirror fermions as a
consequence of the fact that the conjugation operator on fermions gives
independent fields. In addition, there was arbitrariness in the construction
of the potential in the bosonic sector associated with the step of eliminating
the auxiliary fields.

\subsection{Real structures on spectral triples}
\label{sect:st}
The first breakthrough came in 1995 with the publication
of Alain Connes' paper ``Noncommutative geometry and reality'' \cite{C95}. In this paper,
the notion of real structure is introduced, motivated by Atiyah's KR theory
and Tomita's involution operator $J.$ A hint for the necessity of the reality
operator can be taken from physics. We have seen that space-time spinors,
which are elements of the Hilbert space satisfy a chirality condition. The
charge conjugation operator, when acting on these spinors, produces a
conjugate element, which in general is independent. It is possible to replace
the chirality condition, with a reality one, known as the Majorana condition
which equates the two. Imposing both conditions, chirality and reality,
simultaneously can only occur in certain dimensions. The action of the
anti-linear isometry $J$ on the algebra $\mathcal{A}$ satisfies the
commutation relation $\left[  a,b^{\mathrm{o}}\right]  =0,$ $\forall
a,b\in\mathcal{A}$ where%
\begin{equation}
b^{\mathrm{o}}=Jb^{\ast}J^{-1},\qquad\forall b\in\mathcal{A}%
\end{equation}
so that  $b^{\mathrm{o}}\in$ $\mathcal{A}^{\mathrm{o}}.$ This gives a
bimodule, using the representation of $\mathcal{A}\otimes\mathcal{A}%
^{\mathrm{o}}$, given by
\begin{equation}
a\otimes b^{\mathrm{o}}\rightarrow aJb^{\ast}J^{-1},\qquad\forall
a,b\in\mathcal{A}%
\end{equation}
We define the fundamental class $\mu$ of the noncommutative space as a class
in the $KR$-homology of the algebra $\mathcal{A}\otimes\mathcal{A}%
^{\mathrm{o}}$ having the involution
\begin{equation}
\tau\left(  a\otimes b^{\mathrm{o}}\right)  =b^{\ast}\otimes\left(  a^{\ast
}\right)  ^{\mathrm{o}},\qquad\forall a,b\in\mathcal{A}%
\end{equation}
The $KR$-homology cycle  implements the involution $\tau$ given by
\begin{equation}
\tau\left(  w\right)  =JwJ^{-1},\qquad\forall w\in\mathcal{A}\otimes
\mathcal{A}^{\mathrm{o}}%
\end{equation}
These imply that the  $KR$-homology is periodic with period $8$ and the
dimension $n$ modulo $8$ is determined from the commutation rules
\begin{equation}
J^{2}=\varepsilon,\qquad JD=\varepsilon^{\prime}DJ,\qquad J\gamma
=\varepsilon^{\prime\prime}\gamma J
\end{equation}
where $\varepsilon,$ $\varepsilon^{\prime},$ $\varepsilon^{\prime\prime}%
\in\left\{  -1,1\right\}  $ are given as function of $n$ modulo $8$ according
to the table%
\begin{equation}%
\begin{tabular}
[c]{l|llllllll}%
$n$ & $0$ & $1$ & $2$ & $3$ & $4$ & $5$ & $6$ & $7$\\
\hline
$\varepsilon$ & $1$ & $1$ & $-1$ & $-1$ & $-1$ & $-1$ & $1$ & $1$\\
$\varepsilon^{\prime}$ & $1$ & $-1$ & $1$ & $1$ & $1$ & $-1$ & $1$ & $1$\\
$\varepsilon^{\prime\prime}$ & $1$ &  & $-1$ &  & $1$ &  & $-1$ &
\end{tabular}
\end{equation}
It is not surprising that this table agrees with the one obtained by
classifying in which dimensions a spinor obey the Majorana and Weyl
conditions. The intersection form $K_{\ast}\left(  \mathcal{A}\right)  \times
K_{\ast}\left(  \mathcal{A}\right)  \rightarrow\mathbb{Z}$ is obtained from
the Fredholm index of $D$ in $K_{\ast}\left(  \mathcal{A}\otimes
\mathcal{A}^{\mathrm{o}}\right)  .$ Using the Kasparov intersection product,
Poincare duality is formulated in terms of the invertibility of $\mu$ and that
$D$ is an operator of order one implies the condition%
\begin{equation}
\left[  \left[  D,a\right]  ,b^{\mathrm{o}}\right]  =0,\qquad\forall
a,b\in\mathcal{A}%
\end{equation}
Next we consider automorphisms of the algebra $\mathcal{A}$ denoted by
$\mathrm{Aut}\left(  \mathcal{A}\right)  .$ This comprises both of inner and
outer automorphisms. Inner automorphisms $\mathrm{Int}\left(  \mathcal{A}%
\right)  $ is a normal subgroup of $\mathrm{Aut}\left(  \mathcal{A}\right)  $
defined by
\begin{equation}
\alpha\left(  f\right)  =ufu^{\ast},\qquad\forall f\in\mathcal{A},\qquad
u\,u^{\ast}=u^{\ast}u=1
\end{equation}
The group $\mathrm{Aut}^{+}\left(  \mathcal{A}\right)  $ of automorphisms of
the involutive algebra $\mathcal{A}$ are implemented by a unitary operator $U$
in $\mathcal{H}$ commuting with $J$ satisfying
\begin{equation}
\alpha\left(  x\right)  =UxU^{-1}\,\,\qquad\forall x\in\mathcal{A}%
\end{equation}
For Riemannian manifolds $M$, this plays the role of the group of
diffeomorphisms $\mathrm{Diff}^{+}\left(  M\right)  ,$ which preserves the
$K$-homology fundamental class of $M.$ Let $\mathcal{E}$ be a finite
projective, hermitian right $\mathcal{A}$-module, and define the algebra
$\mathcal{B}=\mathrm{End}\left(  \mathcal{A}\right)  $ as the Morita
equivalence of the algebra $\mathcal{A}$ with a hermitian connection $\nabla$
on $\mathcal{E}$ defined as the linear map $\nabla:\mathcal{E\rightarrow
E\otimes}_{\mathcal{A}}\Omega_{D}^{1}$ satisfying
\begin{align*}
\nabla\left(  \zeta a\right)    & =\left(  \nabla\zeta\right)  a+\zeta\otimes
da,\qquad\forall\zeta\in\mathcal{E},\,a\in\mathcal{A}\\
d\left(  \zeta,\eta\right)    & =\left(  \zeta,\nabla\eta\right)  -\left(
\nabla\zeta,\eta\right)  ,\qquad\forall\zeta,\,\eta\in\mathcal{E}%
\end{align*}
where $da=\left[  D,a\right]  $ and $\Omega_{D}^{1}$ is the bimodule of
operators of the form
\begin{equation}
A=%
{\displaystyle\sum\limits_{i}}
a_{i}\left[  D,b_{i}\right]  ,\qquad a_{i},b_{i}\in\mathcal{A}%
\end{equation}
Since any algebra is Morita equivalent to itself with $\mathcal{E}%
=\mathcal{A},$ applying the construction given above yields the inner
deformation of the spectral geometry. The unitary equivalence is implemented
by the representation $u\rightarrow\widetilde{U}=u\left(  Ju\,J^{-1}\right)
=u\left(  u^{\mathrm{o}}\right)  ^{\ast}$ so that the Dirac operator that
includes inner fluctuations%
\begin{equation}
D_{A}=D+A+JAJ^{-1}%
\end{equation}
where $A=A^{\ast}$ transforms as $D_{A}\rightarrow\widetilde{U}D_{A}%
\widetilde{U}^{-1}$ provided that
\begin{equation}
A\rightarrow u\,Au^{\ast}+u\left[  D,u^{\ast}\right]
\end{equation}
This will ensure that the inner product
\begin{equation}
\left(  \Psi,D_{A}\Psi\right)
\end{equation}
is invariant under the transformation $\Psi\rightarrow\widetilde{U}\Psi.$ This
expression will then take care of all fermionic interactions which, as will be
seen in the next section, removes the arbitrariness in specifying the action of
the connection on the Hilbert space.

\subsection{The spectral action principle}
The next breakthrough came a year later in 1996 in the work of Chamseddine and
Connes entitled ``The spectral action principle'' \cite{CC96}. The basic observation is that
for a noncommutative space defined by spectral data, the emphasis is shifted
from the coordinates $x$ of a geometric space to the spectrum $\Sigma
\sqsubset\mathbb{R}$ of the operator $D$. We postulate the following
hypothesis%
\begin{equation}
\text{The physical action depends only on }\Sigma
\end{equation}
The existence of Riemannian manifolds which are isospectral but not isometric
shows that the spectral action principle is stronger than the usual
diffeomorphism invariance. In the usual Riemannian case the group
$\mathrm{Diff}\left(  M\right)  $ of diffeomorphisms of $M$ is canonically
isomorphic to the group $\mathrm{Aut}\left(  \mathcal{A}\right)  $ of
automorphisms of the algebra $\mathcal{A}=C^{\infty}\left(  M\right)  .$ To
each $\varphi\in\mathrm{Diff}\left(  M\right)  $ one associates the algebra
preserving map $\alpha_{\varphi}:\mathcal{A}\rightarrow\mathcal{A}$ given by
\begin{equation}
\alpha_{\varphi}\left(  f\right)  =f\circ\varphi^{-1}\qquad\forall f\in
C^{\infty}\left(  M\right)  =\mathcal{A}%
\end{equation}
The prescription to determine the bosonic action with some cutoff energy scale
$\Lambda$ is to first replace the Hilbert space $\mathcal{H}$ by the subspace
$\mathcal{H}_{\Lambda}$ defined by
\begin{equation}
\mathcal{H}_{\Lambda}=\mathrm{range\,}\chi\left(  \frac{D}{\Lambda}\right)
\end{equation}
where $\chi$ is a suitable smooth positive function, restricting both $D$ and
$\mathcal{A}$ to this subspace maintaining the commutation relations for the
algebra. This procedure is superior to the lattice approximation because it
does respect the geometric symmetry group. The {\em spectal action functional} is then given by the
$$
\tr \chi\left( \frac D \Lambda \right).
$$
For a noncommutative space which is 
a tensor product of a continuous manifold times a discrete space, the functional
$\tr \chi\left(  \frac{D}{\Lambda}\right)  $ can be expanded in an asymptotic series in $\Lambda$, rendering the computation amenable to a heat kernel expansion. This procedure
will be illustrated in the next section. More general methods to analyze the spectral action have also been developed, see \cite{FGLV98} for an early result and also the recent book \cite{EI18}. An interpretation of the spectral action as the von Neumann entropy of a second-quantized spectral triple has been found recently in \cite{CCS18} ({\em cf.} \cite{DK19}).

To summarize, the breakthroughs carried out in the short period 1995-1996,
defining the reality operator $J$ and developing the spectral action principle
will allow to remove the ambiguities encountered before in the construction of
the noncommutative spectral Standard Model.

\section{The spectral Standard Model}

At the time that the spectral action was formulated, it was clear that this principle forms a unifying framework for gravity and particle physics of the Standard Model. As said, this led to much activity ({\em cf.} \cite{SUW02}) in the years that followed. Also shortcomings of the approach were pointed out quite quickly, such as the notorious fermion-doubling problem \cite{LMMS97,GIS98}. This doubling ---or actually, quadrupling--- was due to the incorporation of left-right, particle-anti-particle degrees of freedom both in the continuum spinor space and in the finite noncommutative space. At the technical level this was a crucial starting point, allowing for a product geometry to describe gravity coupled to the Standard Model.

Nevertheless, it was a somewhat disturbing feature which, together with the apparent arbitrariness of the choice of a finite geometry and the abscence of neutrino mixing in the model, led Connes to eventually resolve these problems in \cite{C06}. At the same time John Barrett \cite{Bar06} arrived at the same conclusion (see also the recent uniqueness result \cite{Bes19}), even though his motivation came from the desire to have noncommutative geometry with a Lorentzian signature. 

The crucial insight in both of these works is that one should allow for a KO-dimension for the finite space $F$ which is different from the metric dimension (which is zero). More specifically, the KO-dimension of the finite space should be 6 (modulo 8), so that the product of the continuum $M$ with $F$ is 10 modulo 8. The precise structure of the spectral Standard Model (see Section \ref{sect:spectr-SM}) is then best understood using the classification of all irreducible finite noncommutative geometries of KO-dimension 6 which we now briefly recall.

\subsection{Classification of irreducible geometries}
\label{sect:irr}
In \cite{CC07b} Chamseddine and Connes classified {\it irreducible} finite real spectral triples of KO-dimension 6. This lead to a remarkably concise list of spectral triples, based on the matrix algebras $M_N(\C) \oplus M_N(\C)$ for some $N$. We remark that earlier classification results were obtained \cite{Kra97,PS98} which were also exploited in a search Beyond the Standard Model (see Remark \ref{rem:beyond-sm} below).



\begin{defn}
A finite real spectral triple $(A,H,D;J, \gamma)$ is called {\em irreducible} if the triple $(A,H,J)$ is irreducible. More precisely, we demand that:
\begin{enumerate}
\item The representations of $A$ and $J$ in $H$ are irreducible;
\item The action of $A$ on $H$ has a separating vector.
\end{enumerate}
\end{defn}
We will prove the main result of \cite{CC07b} using an alternative approach which is based on \cite[Sect. 3.4]{Sui14}.

\begin{thm}
\label{thm:irr-geom}
Let $(A,H,D;J,\gamma)$ be an irreducible finite real spectral triple of KO-dimension 6. Then there exists a positive integer $N$ such that $A \simeq M_N(\C) \oplus M_N(\C)$. 
\end{thm}
\proof
Let $(A,H,D;J,\gamma)$ be  an arbitrary finite real spectral triple. We may then decompose
$$
A= \bigoplus_{i=1}^N M_{n_i}(\C), \qquad H = \bigoplus_{i,j=1}^N \C^{n_i} \otimes( \C^{n_j})^\circ \otimes V_{ij},
$$
with $V_{ij}$ corresponding to the multiplicities as before. Now each $\C^{n_i} \otimes \C^{n_j}$ is an irreducible representation of $A$, but in order for $H$ to support a real structure $J:H \to H$ we need both $\C^{n_i} \otimes (\C^{n_j})^\circ$ and $\C^{n_j} \otimes (\C^{n_i})^\circ$ to be present in $H$. Moreover, an old result of Wigner \cite{Wig60} for an anti-unitary operator with $J^2 =1$ assures that already with multiplicities $\dim V_{ij}=1$ there exists such a real structure. Hence, the irreducibility condition (1) above yields 
$$
H = \C^{n_i} \otimes (\C^{n_j})^\circ \oplus  \C^{n_j} \otimes (\C^{n_i})^\circ,
$$
for some $i,j \in \{ 1,\ldots, N\}$. 
Then, let us consider condition (2) on the existence of a separating vector. Note first that the representation of $A$ in $H$ is faithful only if $A= M_{n_i}(\C) \oplus M_{n_j}(\C)$. Second, the stronger condition of a separating vector $\xi$ then implies $n_i = n_j$, as it is equivalent to $A' \xi = H$ for the commutant $A'$ of $A$ in $H$. Namely, since $A' =  M_{n_j}(\C) \oplus M_{n_i}(\C)$ with $\dim A' = n_i^2 + n_j^2$, and $\dim H = 2n_i n_j$ we find the desired equality $n_i=n_j$. 
\endproof

With the complex finite-dimensional algebras $A$ given as a direct sum $M_N(\C) \oplus M_N(\C)$,\footnote{The case $N=1$ was exploited successfully in \cite{DS11} for a noncommutative description of abelian gauge theories.} the additional demand that $H$ carries a symplectic structure $I^2=-1$ yields real algebras of which $A$ is the complexification. We see that this requires $N=2k$ so that one naturally considers triples $(A,H,J)$ for which
\begin{equation}
\label{eq:classif}
A= M_{k}(\bH) \oplus M_{2k}(\C); \qquad H= \C^{2(2k)^2}.
\end{equation}

\subsection{Noncommutative geometry of the Standard Model} 
\label{sect:spectr-SM}

The above classification of irreducible finite geometries of KO-dimension 6 forms the starting point for the derivation of the Standard Model from a noncommutative manifold \cite{CCM07}. Hence, it is based on the matrix algebra $M_N(\C) \oplus M_N(\C)$ for $N \geq 1 $. Let us make the following two additional requirements on the irreducible finite geometry $(A,H_F,D_F;J_F,\gamma_F)$:
\begin{enumerate}
\item The finite-dimensional Hilbert space $H_F$ carries a symplectic structure $I^2 = -1$;
\item the grading $\gamma_F$ induces a non-trivial grading on $A$, by mapping
$$
a \mapsto \gamma_F a \gamma_F,
$$
and selects an even subalgebra $A^\ev \subset A$ consisting of elements that commute with $\gamma_F$.
\end{enumerate}
But the first demand sets $A=M_k(\bH) \oplus M_{2k}(\C)$, represented on the Hilbert space $\C^{2(2k)^2}$. The second requirement sets $k \geq 2$; we will take the simplest $k=2$ so that $H_F = \C^{32}$.
\footnote{Also other algebras that appear in the classification of irreducible geometries of KO-dimension have been considered in the literature: besides the case $N=4$ that we consider here the simplest case $N=1$ is relevant for the noncommutative geometric description of quantum electrodynamics \cite{DS11} and the case $N = 8$ leads to the `grand algebra' of \cite{DLM14,DLM14b}. }
 Indeed, this allows for a $\gamma_F$ such that 
\begin{align}
A^\ev &= \bH_R \oplus \bH_L \oplus M_4(\C), \nn
\intertext{where $\bH_R$ and $\bH_L$ are two copies (referred to as {\em right} and {\em left}) of the quaternions; they are the diagonal of $M_2(\bH) \subset A$. The Hilbert space can then be decomposed according to the defining representations of $A^\ev$, }
\label{eq:HF-PS}
H_F &= (\C^2_R \oplus \C^2_L) \otimes (\C^4)^{\circ} \oplus \C^4 \otimes ( (\C^2_R)^{\circ} \oplus (\C^2_L)^{\circ}).
\intertext{According to this direct sum decomposition, we write }
\label{eq:dirac-sm}
D_F&= \begin{pmatrix} {S}&{T^*}\\ {T}&{\bar S} \end{pmatrix}
\end{align}
Moreover, $J_F$ is the anti-unitary operator that flips the two 16-dimensional components in Equation \eqref{eq:HF-PS}. 

The key result is that if we assume that $T$ is non-trivial, then the first-order condition selects the maximal subalgebra of the Standard Model, that is to say, $A_F= \C \oplus \bH \oplus M_3(\C)$.

\begin{prop}{\cite[Prop. 2.11]{CCM07}}
\label{prop:subalg-sm}
Up to $*$-automorphisms of $A^\ev$, there is a unique $*$-subalgebra $A_F\subset A^\ev$ of maximal dimension that allows $T \neq 0$ in \eqref{eq:dirac-sm}. It is given by
$$
A_F = \left\{ \left( q_\lambda, q , \begin{pmatrix} q &0 \\ 0&m \end{pmatrix} \right): \lambda \in \C, q \in \bH_L, m \in M_3(\C)  \right\} \subset \bH_R \oplus \bH_L \oplus M_4(\C),
$$
where $\lambda \mapsto q_\lambda$ is the embedding of $\C$ into $\bH$, with
$$
q_\lambda = \begin{pmatrix} \lambda & 0 \\ 0 & \bar\lambda \end{pmatrix}.
$$
Consequently, $A_F \simeq \C \oplus \bH \oplus M_3(\C)$. 
\end{prop}
The restriction of the representation of $A$ on $H_F$ to the subalgebra $A_F$ gives a decomposition of $H_F$ into irreducible (left and right) representations of $\C$, $\H_L$ and $M_3(\C)$. For instance,
\begin{equation}
\label{eq:decomp-subalg}
 (\C^2_R \oplus \C^2_L) \otimes (\C^4)^{\circ} \leadsto (\C \oplus \overline \C \oplus \C^2_L) \otimes \left ((\C)^\circ \oplus (\C^3)^\circ \right).
\end{equation}
and similarly for $\C^4 \otimes ( (\C^2_R)^{\circ} \oplus (\C^2_L)^{\circ})$. In order to connect to the physics of the Standard Model, let us introduce an orthonormal basis for $H_F$ that can be recognized as the fermionic particle content of the Standard Model, and subsequently write the representation of $A_F$ in terms of this basis. 

We let the subspace of $H_F$ displayed in Equation \eqref{eq:decomp-subalg} be represented by basis vectors $\{ \nu_R ,e_R, (\nu_L,e_L)\}$ of the so-called {\em lepton space} $H_l$ and basis vectors $\{ u_R,d_R,(u_L,d_L)\}$ of the {\em quark space} $H_q$. Their reflections with respect to $J_F$ are the {\em anti-lepton space} $H_{\bar l}$ and the {\em anti-quark space} $H_{\bar q}$, spanned by $\{ \bar{\nu_R} ,\bar{e_R}, (\bar{\nu_L},\bar{e_L})\}$ and $\{\bar{u_R}, \bar{d_R},( \bar{u_L},\bar{d_L})\}$, respectively. The three colors of the quarks are given by a tensor factor $\C^3$ and when we take into account {\em three generations} of fermions and anti-fermions by tripling the above finite-dimensional Hilbert space we obtain
\begin{align*}
H_F := \left( H_l \oplus H_{\bar l} \oplus H_q \oplus H_{\bar q} \right)^{\oplus3} .
\end{align*}
Note that $H_l = \C^4$, $H_q=\C^4 \otimes \C^3$, $H_{\bar l} = \C^4$, and $H_{\bar q} = \C^4 \otimes \C^3$. 

An element $a=(\lambda,q,m)\in A_F$ acts on the space of leptons $H_l$ as $q_\lambda \oplus q$, and acts on the space of quarks $H_q$ as $(q_\lambda \oplus q) \otimes 1_3$. 
For the action of $a$ on an anti-lepton $\bar l\in H_{\bar l}$ we have $a\bar l = \lambda 1_4\bar l$, and on an anti-quark $\bar q\in H_{\bar q}$ we have $a\bar q = (1_4 \otimes m) \bar q$.

The $\Z_2$-grading $\gamma_F$ is such that left-handed particles have eigenvalue $+1$ and right-handed particles have eigenvalue $-1$. The anti-linear operator $J_F$ interchanges particles with their anti-particles, so $J_F f = \bar f$ and $J_F \bar f = f$, with $f$ a lepton or quark. 

\bigskip

The first indication that the subalgebra $A_F$ is relevant for the Standard Model ---to say the least--- comes from the fact that the Standard Model gauge group can be derived from the unitaries in $A_F$. We restrict to the {\em unimodular gauge group},
$$
\mathrm{SU}(A_F) = \left\{ u \in A_F: u^* u = uu^* = 1, \det(u) = 1\right\}
$$
where $\det$ is the determinant of the action of $u$ in $H_F$. It then follows that, up to a finite abelian group we have 
$$
\mathrm{SU}(A_F) \sim U(1) \times SU(2) \times SU(3)
$$
and the hypercharges are derived from the unimodularity condition to be the usual ones:
\begin{align*}
\begin{array}{l|cccccccc}
\text{Particle} & \nu_R & e_R & \nu_L & e_L & u_R & d_R & u_L & d_L \\
\hline
\text{Hypercharge} & 0 & -2 & -1 & -1 & \frac43 & -\frac23 & \frac13 & \frac13 \\
\end{array}
\end{align*}

\bigskip

Let us now turn to the form of the finite Dirac operator, and see what we can say about the components of the matrix $D_F$ as displayed in \eqref{eq:dirac-sm}. Recall that we are looking for a self-adjoint operator $D_F$ in $H_F$ that commutes with $J_F$, anti-commutes with $\gamma_F$, and fulfills the first-order conditions with resepct to $A_F$:
$$
[[D,a],J b J^{-1}]=0; \qquad (a,b \in A_F).
$$
We also require that $D_F$ commutes with the subalgebra $\C_F = \{ (\lambda,\lambda,0) \} \subset A_F$ which physically speaking corresponds to the fact that the photon remains massless. Then it turns out \cite[Theorem 1]{C06} (see also \cite[Theorem 2.21]{CCM07}\label{page:moduli-dirac}) that any $D_F$ that satisfies these assumptions is of the following form: in terms of the decomposition of $H_F$ in particle ($H_l^{\oplus 3} \oplus H_q^{\oplus 3}$) and anti-particles ($H_{\bar l}^{\oplus 3} \oplus H_{\bar q}^{\oplus 3}$) the operator $S$ is
\begin{align}
  S_l &:= \left.S\right|_{H_l^{\oplus 3}} = \begin{pmatrix} 0&0&Y_\nu^*&0 \\0&0&0&Y_e^*\\ Y_\nu&0&0&0\\ 0&Y_e&0&0\end{pmatrix} ,
    \label{eq:yukawa-l} \\
 S_q \otimes 1_3 &:= \left.S\right|_{H_q^{\oplus 3}} =\begin{pmatrix} 0&0&Y_u^*&0 \\0&0&0&Y_d^* \\Y_u&0&0&0 \\0&Y_d&0&0\end{pmatrix} \otimes1_3 ,\label{eq:yukawa-q}
\end{align}
where $Y_\nu$, $Y_e$, $Y_u$ and $Y_d$ are some $3\times3$ matrices acting on the three generations, and $1_3$ acting on the three colors of the quarks. The symmetric operator $T$ only acts on the right-handed (anti)neutrinos, so it is given by $T\nu_R = Y_R\bar{\nu_R}$, for a certain $3\times3$ symmetric matrix $Y_R$, and $Tf=0$ for all other fermions $f\neq\nu_R$. Note that $\nu_R$ here stands for a vector with $3$ components for the number of generations.

The above classification result shows that the Dirac operators $D_F$ give all the required features, such as mixing matrices for quarks and leptons, unbroken color and the see-saw mechanism for right-handed neutrinos. Let us illustrate the latter in some more detail. The mass matrix restricted to the subspace of $H_F$ with basis $\{\nu_L, \nu_R, \bar{\nu_L}, \bar{\nu_R}\}$ is given by
\begin{align*}
\begin{pmatrix} 0&Y_\nu^*&Y_R^*&0 \\ Y_\nu&0&0&0\\ Y_R&0&0&\bar Y_\nu^* \\ 0&0&\bar Y_\nu&0\end{pmatrix} .
\end{align*}
Suppose we consider only one generation, so that $Y_\mu = m_\nu$ and $Y_R = m_R$ are just scalars. The eigenvalues of the above mass matrix are then given by
\begin{align*}
\pm \frac12 m_R \pm \frac12 \sqrt{{m_R}^2 + 4{m_\nu}^2} .
\end{align*} 
If we assume that $m_\nu \ll m_R$, then these eigenvalues are approximated by $\pm m_R$ and $\pm\frac{{m_\nu}^2}{m_R}$. This means that there is a heavy neutrino, for which the Dirac mass $m_\nu$ may be neglected, so that its mass is given by the Majorana mass $m_R$. However, there is also a light neutrino, for which the Dirac and Majorana terms conspire to yield a mass $\frac{{m_\nu}^2}{m_R}$, which is in fact much smaller than the Dirac mass $m_\nu$. This is called the \emph{seesaw mechanism}. Thus, even though the observed masses for these neutrinos may be very small, they might still have large Dirac masses (or Yukawa couplings).

\begin{rem}
Of course, in the physical applications one chooses $Y_\nu, Y_e$ to be the {\em Yukawa mass matrices} and $Y_R$ is the {\em Majorana mass matrix}. There has been searches for additional conditions to be satisfied by the spectral triple $(A_F,H_F,D_F)$ to further constrain the form of $D_F$, see for instance \cite{BBB15,BF18,KL18,DAS18,DS18}. 
\end{rem}

\subsection{The gauge and scalar fields as inner fluctuations}
We here derive the precise form of internal fluctuations $A_\mu$ for the above spectral triple of the Standard Model (following \cite[Sect. 3.5]{CCM07} or \cite[Sect. 11.5]{Sui14}).

Take two elements $a=(\lambda,q,m)$ and $b=(\lambda',q',m')$ of the algebra $\A = C^\infty(\C\oplus\bH \oplus M_3(\C))$. According to the representation of $A_F$ on $H_F$, the inner fluctuations $A_\mu = -ia\partial_\mu b$ decompose as 
\begin{align*}
\Lambda_\mu &:= -i\lambda\partial_\mu\lambda'; \qquad
\Lambda_\mu' := -i\bar\lambda\partial_\mu\bar\lambda'
\intertext{on $\nu_R$ and $e_R$, respectively, and as } 
Q_\mu &:= -iq\partial_\mu q';\qquad
V_\mu' := -im\partial_\mu m'
\end{align*} acting on $(\nu_l,e_L)$ and $H_{\bar q}$, respectively. On all other components of $H_F$ the gauge field $A_\mu$ acts as zero. Imposing the hermiticity $\Lambda_\mu=\Lambda_\mu^*$ implies $\Lambda_\mu\in\R$, and also automatically yields $\Lambda_\mu' = -\Lambda_\mu$. Furthermore, $Q_\mu = Q_\mu^*$ implies that $Q_\mu$ is a real-linear combination of the Pauli matrices, which span $i\,su(2)$. Finally, the condition that $V_\mu'$ be hermitian yields $V_\mu' \in i\,u(3)$, so $V_\mu'$ is a $U(3)$ gauge field. As mentioned above, we need to impose the unimodularity condition to obtain an $SU(3)$ gauge field. Hence, we require that the trace of the gauge field $A_\mu$ over $H_F$ vanishes, and we obtain
\begin{align*}
\left.\tr\right|_{H_{\bar l}}\big( \Lambda_\mu 1_4 \big) + \left.\tr\right|_{H_{\bar q}}\big( 1_4\otimes V_\mu' \big) = 0  \quad\Longrightarrow\quad  \tr(V_\mu') = - \Lambda_\mu .
\end{align*}
Therefore, we can define a traceless $SU(3)$ gauge field $V_\mu$ by $\bar V_\mu := - V_\mu' - \frac13 \Lambda_\mu$. 
The action of the gauge field $B_\mu = A_\mu - J_FA_\mu J_F^{-1}$ on the fermions is then given by
\begin{align}
\left.B_\mu\right|_{H_l} &= \begin{pmatrix} 0&0&  \\ 0&-2\Lambda_\mu&  \\ &&Q_\mu-\Lambda_\mu1_2\end{pmatrix} , \notag\\
\label{eq:Gauge_field_SM}
\left.B_\mu\right|_{H_q} &= \begin{pmatrix} \frac43\Lambda_\mu1_3+V_\mu&0&  \\ 0&-\frac23\Lambda_\mu1_3+V_\mu& \\ &&(Q_\mu+\frac13\Lambda_\mu1_2)\otimes1_3+1_2\otimes V_\mu\end{pmatrix} .
\end{align}
for some $U(1)$ gauge field $\Lambda_\mu$, an $SU(2)$ gauge field $Q_\mu$ and an $SU(3)$ gauge field $V_\mu$.

Note that the coefficients in front of $\Lambda_\mu$ in the above formulas are precisely the aforementioned (and correct!) hypercharges of the corresponding particles. 

\bigskip

Next, let us turn to the scalar field $\phi$, which is given by
\begin{align}
\label{eq:higgs_field_SM}
\left.\phi\right|_{H_l} &= \mattwo{0}{Y^*}{Y}{0} , & \left.\phi\right|_{H_q} &= \mattwo{0}{X^*}{X}{0} \otimes1_3 , & \left.\phi\right|_{H_{\bar l}} &= 0 , & \left.\phi\right|_{H_{\bar q}} &= 0 ,
\end{align}
where we now have, for complex fields $\phi_1,\phi_2$, 
\begin{align*}
Y &= \mattwo{Y_\nu\phi_1}{-Y_e\bar\phi_2}{Y_\nu\phi_2}{Y_e\bar\phi_1} , & X &= \mattwo{Y_u\phi_1}{-Y_d\bar\phi_2}{Y_u\phi_2}{Y_d\bar\phi_1} . 
\end{align*}
The scalar field $\Phi$ is then given by
\begin{align}
\label{eq:Higgs_field_SM}
\Phi = D_F + \mattwo{\phi}{0}{0}{0} + J_F\mattwo{\phi}{0}{0}{0}J_F^* = \mattwo{S+\phi}{T^*}{T}{\bar{(S+\phi)}} .
\end{align}
Finally, one can compute that the action of the gauge group $\mathrm{SU}(A_F)$ by conjugation on the fluctuated Dirac operator
\begin{align*}
D_\omega  = \dirac\otimes 1 + \gamma^\mu\otimes B_\mu + \gamma_M\otimes\Phi
\end{align*}
is implemented by
\begin{gather*}
\Lambda_\mu \mapsto \Lambda_\mu - i \lambda\partial_\mu\bar\lambda , \quad
Q_\mu \mapsto qQ_\mu q^* - iq\partial_\mu q^* , \quad
\bar V_\mu \mapsto m\bar V_\mu m^* - im\partial_\mu m^* , \\
H \mapsto \bar\lambda\,q H ,
\end{gather*}
for $\lambda\in C^\infty\big(M,U(1)\big)$, $q\in C^\infty\big(M,SU(2)\big)$ and $m\in C^\infty\big(M,SU(3)\big)$ and we have written the {\em Higgs doublet} as
$$
 H:= \vectwo{\phi_1+1}{\phi_2}
$$
For the detailed computation we refer to \cite[Sect. 3.5]{CCM07} or \cite[Prop. 11.5]{Sui14}. 

\bigskip

Summarizing, the gauge fields derived take values in the Lie algebra $u(1) \oplus su(2) \oplus su(3)$ and transform according to the usual Standard Model gauge transformations. The scalar field $\phi$ transforms as the Standard Model Higgs field in the defining representation of $SU(2)$, with hypercharge $-1$. 

\subsection{Spectral action}
The spectral action for the above spectral Standard Model has been computed in full detail in \cite[Section 4.2]{CCM07} and confirmed in {\em e.g.} \cite[Theorem 11.10]{Sui14}. Since it would lie beyond the scope of the present review, we refrain from repeating this computation. Instead, we summarize the main result, which is that the Lagrangian derived from the spectral action is 
\begin{align*}
S_B= \int &\Bigg( \frac{48\chi_4\Lambda^4}{\pi^2} - \frac{c\chi_2\Lambda^2}{\pi^2} + \frac{d\chi(0)}{4\pi^2} + \left(\frac{c\chi(0)}{24\pi^2} - \frac{4\chi_2\Lambda^2}{\pi^2} \right) s - \frac{3\chi(0)}{10\pi^2} (C_{\mu\nu\rho\sigma})^2 
 \notag\\
&\quad+ \frac14 Y_{\mu\nu} Y^{\mu\nu} + \frac14 W_{\mu\nu}^a W^{\mu\nu,a} + \frac14 G_{\mu\nu}^i G^{\mu\nu,i} + \frac{b\pi^2}{2a^2\chi(0)} |H|^4 \notag\\
&\quad- \frac{2a\chi_2\Lambda^2 - e\chi(0)}{a\chi(0)} |H|^2 + \frac{1}{12} s |H|^2 + \frac12 |D_\mu H|^2 \Bigg) \sqrt{g} d^4x ,
\end{align*}
where $\chi_j = \int_0^\infty \chi(v) v^{j-1} dv$ are the moments of the function $\chi$, $j>0$, $s=-R$ is the scalar curvature, $Y_{\mu\nu}, W_{\mu\nu}$ and $G_{\mu\nu}$ are the field strengths of $Y_\mu, Q_\mu$ and $V_\mu$, respectively and the covariant derivative $D_\mu H$ is given by
\begin{align}
\label{eq:Higgs_kin_gauge}
D_\mu H = \partial_\mu H + \frac12 i g_2 W_\mu^a \sigma^a H - \frac12 i g_1 Y_\mu H .
\end{align}
Moreover, we have defined the following constants
\begin{align}
\label{eq:abcde_SM}
a &= \tr\big(Y_\nu^*Y_\nu + Y_e^*Y_e + 3Y_u^*Y_u + 3Y_d^*Y_d\big) , \notag\\
b &= \tr\big((Y_\nu^*Y_\nu)^2 + (Y_e^*Y_e)^2 + 3(Y_u^*Y_u)^2 + 3(Y_d^*Y_d)^2\big) , \notag\\
c &= \tr\big(Y_R^*Y_R\big) , \\
d &= \tr\big((Y_R^*Y_R)^2\big) , \notag\\
e &= \tr\big(Y_R^*Y_R Y_\nu^*Y_\nu\big) . \notag
\end{align}
The normalization of the kinetic terms imposes a relation between the coupling constants $g_1,g_2,g_3$ and the coefficients $\chi_0$, of the form 
\begin{align}
\label{eq:couplings_norm}
\frac{\chi(0)}{2\pi^2} {g_3}^2 = \frac{\chi(0)}{2\pi^2} {g_2}^2 = \frac{5\chi(0)}{6\pi^2} {g_1}^2 = \frac14 .
\end{align}
The coupling constants are then related by 
\begin{align*}
{g_3}^2 = {g_2}^2 = \frac53 {g_1}^2 ,
\end{align*}
which is precisely the relation between the coupling constants at unification, common to grand unified theories (GUT). We shall further discuss this in Section \ref{sect:pheno}. 

\subsection{Fermionic action in KO-dimension 6}
As already announced above, the shift to KO-dimension 6 for the finite space solved the fermion doubling problem of \cite{LMMS97}. Let us briefly explain how this works, following \cite{C06}.
  
The crucial observation is that in KO-dimension $2 \equiv 4+6 \mod 8$ the following pairing 
$$
( \psi, \psi') \mapsto (J \psi, D_\omega \psi')
$$
is a skew-symmetric form on the $+1$-eigenspace of $\gamma$ in $\H$. This skew-symmetry is in concordance with the Grassmann nature of fermionic fields $\psi$, guaranteeing that the following action functional is in fact non-zero:
$$
S_F = \frac 12 \langle J \xi , D_A \xi \rangle
$$
for $\xi$ a Grassmann variable in the $+1$-eigenspace of $\gamma$. 

This then solves the fermion doubling, or actually quadrupling as follows. First, the restriction to the chiral subspace of $\gamma$ takes care of a factor of two. Then, the functional integral involving anti-commuting Grassman variables delivers a Pfaffian, which takes care of a square root. That this indeed works has been worked out in full detail for the case of the Standard Model in \cite[Section 4.4.1]{CCM07} or \cite[Section 11.4]{Sui14}.

\subsection{Phenomenological consequences}
\label{sect:pheno}

The first phenomenological consequence one can derive from the spectral Standard Model is an upper bound on the mass of the top quark. In fact, the appearance of the constant $a$ in both the fermionic and the bosonic action allows to derive 
\begin{align}
\label{eq:masses_ferm_W}
\tr\big(m_\nu^*m_\nu + m_e^*m_e + 3m_u^*m_u + 3m_d^*m_d\big) = 2{g_2}^2{v}^2 = 8 {M_W}^2 .
\end{align}
It is natural to assume that the mass $m_{\text{top}}$ of the top quark is much larger than all other fermion masses, except possibly a Dirac mass that arises from the seesaw mechanism as was described above. If we write $m_\nu = \rho m_{\text{top}}$ then the above relation would yield the constraint
\begin{align}
\label{eq:top_mass}
m_{\text{top}} \lesssim \sqrt\frac8{3+\rho^2} M_W .
\end{align}
The relations \eqref{eq:couplings_norm} between the coupling constants and $\chi(0)$ suggests that we have grand unification of the coupling constants. Moreover, from the action functional we see that the quartic Higgs coupling constant $\lambda$ is related to $\chi(0)$ as well via
$$
\lambda = 24 \frac{b}{a^2} g_2^2.
$$
Thus, the spectral Standard Model imposes relations between the coupling constants and bounds on the fermion masses. These relations were used in \cite{CCM07} as input at (or around) grand unification scale $\Lambda_\GUT$, and then run down using one-loop renormalization group equations to 'low energies' where falsifiable predictions were obtained. 

\begin{figure}
\centering
\includegraphics[scale=.5]{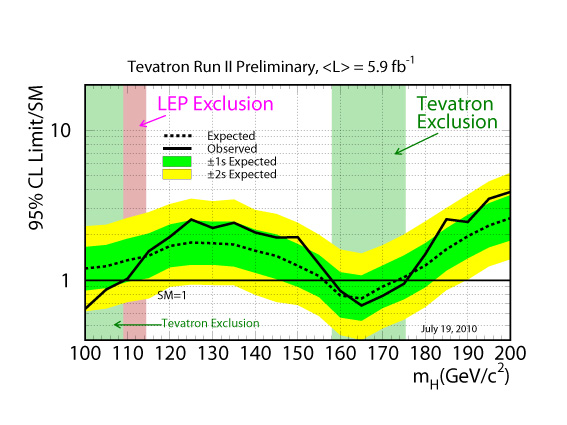}
\caption{Observed and expected exclusion limits for a Standard Model Higgs boson at the 95-percent confidence level for the combined CDF and DZero analyses. (Fermilab) }
\label{fig:fermilab}
\end{figure}

In fact, the mass of the top quark can indeed be found to get an acceptable value, however, for the Higgs mass it was found that 
$$
167 \GeV \leq m_h \leq 176 \GeV.
$$
Given that there were not much models in particle physics around that could produce falsifiable predictions, it is somewhat ironical that the first exclusion results on the mass of the Higgs that appeared in 2009 from Fermilab hit exactly this region. See Figure \ref{fig:fermilab}. And, of course, with the discovery of the Higgs at $m_h  \approx 125.5 \GeV$ in \cite{ATLAS12,CMS12} one could say that the spectral Standard Model was not in a particularly good shape at that time. 

\section{Beyond the Standard Model with noncommutative geometry}

Even though the incompatibility between the spectral Standard Model and the experimental discovery of the Higgs with a relatively low mass was not an easy stroke at the time, it also led to a period of reflection and reconsideration of the premises of the noncommutative geometric approach. In fact, it was the beginning of yet another exciting chapter in our story on the spectral model of gravity coupled with matter. As we will see in this and the next chapter, once again the input from experiment is taken as a guiding principle in our search for the spectral model that goes Beyond the Standard Model.

\begin{rem}
  \label{rem:beyond-sm}
  We do not pretend to give a complete overview of the literature here, but only indicate some of the highlights and actively ongoing research areas.

  Other searches beyond the Standard Model with noncommutative geometry include \cite{ISS04,Ste06,Ste07,Ste09,Ste09b,Ste13}, adopting a slightly different approach to almost-commutative manifolds as we do. 
  
There is another aspect that was studied is the connection between supersymmetry and almost-commutative manifolds. It turned out to be very hard ---if not impossible--- to combine the two. A first approach is \cite{Cha94} and more recently the intersection was studied in \cite{BroS10,BroS11,BBS16}.
  \end{rem}

\subsection{Resilience of the spectral Standard Model}
In 2012 it was realized how a small correction of the spectral Standard Model gives an intriguing possibility to go beyond the Standard Model, solving at the same time a problem with the stability of the Higgs vacuum given the measured low mass $m_h$. This is based on \cite{CC12}, but for which some of the crucial ingredients surprisingly enough were already present in the 2010 paper \cite{CC10}. 

Namely, in the definition of the finite Dirac operator $D_F$ of Equation \ref{eq:dirac-sm}, we can replace $Y_R$ by $Y_R \sigma$, where $\sigma$ is a real scalar field on $M$. Strictly speaking, this brings us out of the class of almost-commutative manifolds $M \times F$, since part of $D_F$ now varies over $M$ and this was the main reason why it was disregarded before. However, since from a physical viewpoint there was no reason to assume $Y_R$ to be constant, it was treated as a scalar field already in \cite{CC10}. This was only fully justified in subsequent papers (as we will see in the next subsections) where the scalar field $\sigma$ arises as the relic of a spontaneous symmetry breaking mechanism, similar to the Higgs field $h$ in the electroweak sector of the Standard Model. We will discuss a few of the existing approaches in the literature in the next few sections. For now, let us simply focus on the phenomenological consequences of this extra scalar field. 

Thus we replace $Y_R$ by $Y_R \sigma$ and analyze the additional terms in the spectral action. The scalar sector becomes
\begin{multline*}
  S_H' := \int_M \bigg( \frac{bf(0)}{2\pi^2} |H|^4 - \frac{2af_2\Lambda^2}{\pi^2} |H|^2  +\frac{ef(0)}{\pi^2} \sigma^2|H|^2\\
-\frac{cf_2\Lambda^2}{\pi^2} \sigma^2 + \frac{df(0)}{4\pi^2} \sigma^4   + \frac{af(0)}{2\pi^2} |D_\mu H|^2 + \frac{1}{4 \pi^2} f(0) c ( \partial_\mu \sigma)^2 \bigg) \sqrt{g} dx,
\end{multline*}
where we ignored the coupling to the scalar curvature.

We exploit the approximation that $m\Sub{top}$, $m_\nu$ and $m_R$ are the dominant mass terms. Moreover, as before we write $m_\nu = \rho m\Sub{top}$. That is, the expressions for $a,b,c,d$ and $e$ in \eqref{eq:abcde_SM} now become
\begin{align*}
a  &\approx m\Sub{top}^2 (\rho^2 +3),\\
b &\approx m\Sub{top}^4 (\rho^4 +3),\\
c  & \approx m_R^2,\\
d&\approx m_R^4,\\
e&\approx \rho^2 m_R^2 m\Sub{top}^2. 
\end{align*}
In a unitary gauge, where $H = \begin{pmatrix} h \\ 0 \end{pmatrix}$, we arrive at the following potential:
$$
\L\Sub{pot}(h,\sigma) = \frac{1}{24} \lambda_h h^4 + \frac12 \lambda_{h \sigma} h^2 \sigma^2 + \frac14 \lambda_\sigma \sigma^4 - \frac{4 g_2^2}{\pi^2} f_2 \Lambda^2 (h^2 + \sigma^2),
$$
where we have defined coupling constants
\begin{align}
\label{eq:scalar-couplings}
\lambda_h &= 24 \frac{\rho^4 + 3}{(\rho^2+3)^2} g_2^2,&
\lambda_{h \sigma} &= \frac{8 \rho^2}{\rho^2 +3}  g_2^2,&
\lambda_\sigma &= 8 g_2^2.
\end{align}
This potential can be minimized, and if we replace $h$ by $v+h$ and $\sigma$ by $w+ \sigma$, respectively, expanding around a minimum for the terms quadratic in the fields, we obtain:
\begin{align*}
\L\Sub{pot}(v+h,w+\sigma)|\Sub{\text{quadratic}} &= \frac16 v^2 \lambda_h v^2 + 2 vw \lambda_{h \sigma} \sigma h + w^2 \lambda_\sigma \sigma^2 \\
&= \frac12 \begin{pmatrix} h & \sigma \end{pmatrix} M^2 \begin{pmatrix} h \\ \sigma \end{pmatrix},
\end{align*}
where we have defined the mass matrix $M$ by 
$$
M^2 = 2 \begin{pmatrix} \frac16 \lambda_h v^2 & \lambda_{h \sigma} vw \\ \lambda_{h \sigma} vw & \lambda_\sigma w^2 \end{pmatrix}.
$$
This mass matrix can be easily diagonalized, and if we make the natural assumption that $w$ is of the order of $m_R$, while $v$ is of the order of $M_W$, so that $v \ll w$, we find that the two eigenvalues are
\begin{align*}
m_+^2 &\sim 2 \lambda_\sigma w^2 + 2 \frac{\lambda_{h \sigma}^2}{\lambda_\sigma} v^2,\\
m_-^2 &\sim 2 \lambda_h v^2 \left( \frac16 - \frac{\lambda_{h \sigma}^2}{\lambda_h \lambda_\sigma}\right).
\end{align*}
We can now determine the value of these two masses by running the scalar coupling constants $\lambda_h, \lambda_{h \sigma}$ and $\lambda_\sigma$ down to ordinary energy scalar using the renormalization group equations for these couplings that were derived in \cite{GLPR10}, referring to \cite{CC12,Sui14} for full details. 
\begin{figure}[t!]
  \centering
\includegraphics[scale=.25]{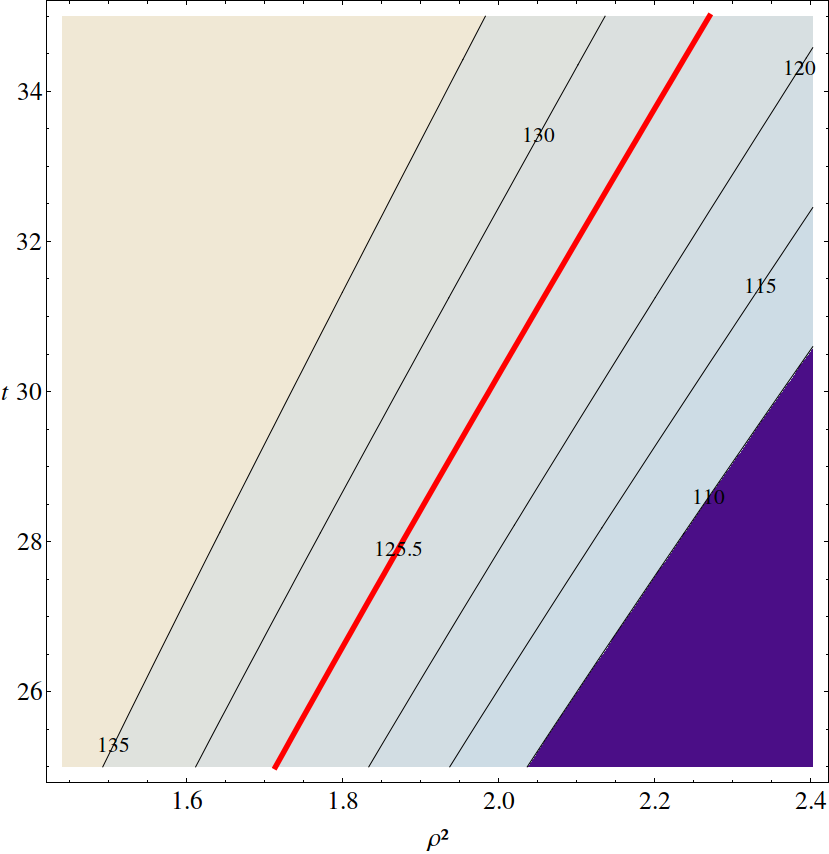}
\caption{A contour plot of the Higgs mass $m_h$ as a function of $\rho^2$ and $t = \log (\Lambda\Sub{GUT}/M_Z)$. The red line corresponds to $m_h = 125.5~\GeV$.}
\label{fig:higgsmass}
\end{figure}
The result varies with the chosen value for $\Lambda_\GUT$ and the parameter $\rho$. The mass of $\sigma$ is essentially given by the largest eigenvalue $m_+$ which is of the order $10^{12}~  \GeV$ for all values of $\Lambda_\GUT$ and the parameter $\rho$. The allowed mass range for the Higgs, {\it i.e.} for $m_-$, is depicted in Figure \ref{fig:higgsmass}. The expected value $m_h=125.5 ~\GeV$ is therefore compatible with the above noncommutative model. Moreover, without the $\sigma$ the $\lambda_h$ turns negative at energies around $10^{12} \GeV$. Furthermore, this calculation implies that there is a relation (given by the red line in the Figure) between the ratio $m_\nu/m\Sub{top}$ and the unification scale $\Lambda_\GUT$.

\subsection{Pati--Salam unification and first-order condition}
\label{sect:patisalam}
In order to see how we one can use the noncommutative geometric approach to go beyond the Standard Model it is important to trace our steps that led to the spectral Standard Model in the previous Section. The route started with the classification of the algebras of the finite space ({\em cf.}  Equation \eqref{eq:classif}). The results show that the only algebras which solve the fermion
doubling problem are of the form $M_{2a}(\mathbb{C})\oplus M_{2a}(\mathbb{C})$
where $a$ is an even integer. An arbitrary symplectic constraint is imposed on
the first algebra restricting it from $M_{2a}(\mathbb{C})$ to $M_{a}%
(\mathbb{H}).$ The first non-trivial algebra one can consider is for $a=2$
with the algebra
\begin{equation}
M_{2}(\mathbb{H})\oplus M_{4}(\mathbb{C}).
\end{equation}
Coincidentally, and as explained in the introduction, the above algebra comes
out as a solution of the two-sided Heisenberg quantization relation between
the Dirac operator $D$ and the two maps from the four spin-manifold and the
two four spheres $S^{4}\times S^{4}$ \cite{CCM14,CCM15}. This removes the
arbitrary symplectic constraint and replaces it with a relation that quantize
the four-volume in terms of two quanta of geometry and have far reaching
consequences on the structure of space-time. We will come back to this in the last Section.

The existence of the chirality operator $\gamma$ that commutes with the
algebra breaks the quaternionic matrices $M_{2}(\mathbb{H})$ to the diagonal
subalgebra and leads us to consider the finite algebra
\begin{equation}
\mathcal{A}_{F}=\mathbb{H}_{R}\oplus\mathbb{H}_{L}\oplus M_{4}(\mathbb{C}).
\end{equation}
This algebras is the simplest candidate to search for new physics beyond the Standard Model. In fact, the inner automorphism group of $\mathcal{A=C}^{\infty}\left(
M\right)  \otimes\mathcal{A}_{F}$ is recognized as the Pati--Salam gauge group $SU(2)_{R}\times SU(2)_{L}\times SU(4)$, and the corresponding gauge bosons appear
as inner perturbations of the (spacetime) Dirac operator \cite{CCS13b}. Thus, we are considering a spectral Pati--Salam model as a candidate beyond the Standard Model. Let us further analyze this model and its phenomenological consequences.

An element of the Hilbert space $\Psi\in\mathcal{H}$ is represented by
\begin{equation}
\Psi_{M}=\left(
\begin{array}
[c]{c}%
\psi_{A}\\
\psi_{A^{^{\prime}}}%
\end{array}
\right)  ,\quad\psi_{A^{\prime}}=\psi_{A}^{c}%
\end{equation}
where $\psi_{A}^{c}$ is the conjugate spinor to $\psi_{A}.$ Thus all primed
indices $A^{\prime}$ correspond to the Hilbert space of conjugate spinors. It
is acted on by both the left algebra $M_{2}\left(  \mathbb{H}\right)  $ and
the right algebra $M_{4}\left(  \mathbb{C}\right)  $. Therefore the index $A$
can take $16$ values and is represented by
\begin{equation}
A=\alpha I
\end{equation}
where the index $\alpha$ is acted on by quaternionic matrices and the index
$I$ \ by $M_{4}\left(  \mathbb{C}\right)  $ matrices. Moreover, when the
grading breaks $M_{2}\left(  \mathbb{H}\right)  $ into $\mathbb{H}_{R}%
\oplus\mathbb{H}_{L}$ the index $\alpha$ is decomposed to $\alpha
=\overset{.}{a},a$ where $\overset{.}{a}=\overset{.}{1},\overset{.}{2}$
(dotted index) is acted on by the first quaternionic algebra $\ \mathbb{H}%
_{R}$ and $a=1,2$ is acted on by the second quaternionic algebra
$\ \mathbb{H}_{L}$. When $M_{4}\left(  \mathbb{C}\right)  $ breaks into
$\mathbb{C}\oplus M_{3}\left(  \mathbb{C}\right)  $ (due to symmetry breaking
or through the use of the order one condition as in \cite{CC07b}) the index
$I$ is decomposed into $I=1,i$ and thus distinguishing leptons and quarks,
where the $1$ is acted on by the $\mathbb{C}$ and the $i$ by $M_{3}\left(
\mathbb{C}\right)  .$ Therefore the various components of the spinor $\psi
_{A}$ are
\begin{align}
\psi_{\alpha I} &  =\left(
\begin{array}
[c]{cccc}%
\nu_{R} & u_{iR} & \nu_{L} & u_{iL}\\
e_{R} & d_{iR} & e_{L} & d_{iL}%
\end{array}
\right)  ,\qquad i=1,2,3\\
&  =\left(  \psi_{\overset{.}{a}1},\psi_{\overset{.}{a}i},\psi_{a1},\psi
_{ai}\right)  ,\qquad a=1,2,\quad\overset{.}{a}=\overset{.}{1},\overset{.}{2}%
\nonumber
\end{align}
This is a general prediction of the spectral construction that there is $16$
fundamental Weyl fermions per family, $4$ leptons and $12$ quarks.

The (finite) Dirac operator can be written in matrix form%
\begin{equation}
D_{F}=\left(
\begin{array}
[c]{cc}%
D_{A}^{B} & D_{A}^{B^{^{\prime}}}\\
D_{A^{^{\prime}}}^{B} & D_{A^{^{\prime}}}^{B^{^{\prime}}}%
\end{array}
\right)  ,\label{eq:dirac}%
\end{equation}
and must satisfy the properties
\begin{equation}
\gamma_{F}D_{F}=-D_{F}\gamma_{F}\qquad J_{F}D_{F}=D_{F}J_{F}%
\end{equation}
where $J_{F}^{2}=1.$ A\ matrix realization of $\gamma_{F}$ and $J_{F}$ are
given by
\begin{equation}
\gamma_{F}=\left(
\begin{array}
[c]{cc}%
G_{F} & 0\\
0 & -\overline{G}_{F}%
\end{array}
\right)  ,\qquad G_{F}=\left(
\begin{array}
[c]{cc}%
1_{2} & 0\\
0 & -1_{2}%
\end{array}
\right)  ,\qquad J_{F}=\left(
\begin{array}
[c]{cc}%
0_{4} & 1_{4}\\
1_{4} & 0_{4}%
\end{array}
\right)  \circ\mathrm{cc}%
\end{equation}
where $\mathrm{cc}$ stands for complex conjugation. These relations, together
with the hermiticity of $D$ imply the relations
\begin{equation}
\left(  D_{F}\right)  _{A^{^{\prime}}}^{B^{^{\prime}}}=\left(  \overline
{D}_{F}\right)  _{A}^{B}\,\qquad\left(  D_{F}\right)  _{A^{^{\prime}}}%
^{B}=\left(  \overline{D}_{F}\right)  _{B}^{A^{\prime}}%
\end{equation}
and have the following zero components \cite{CC10}
\begin{align}
\left(  D_{F}\right)  _{aI}^{bJ} &  =0=\left(  D_{F}\right)  _{\overset{.}{a}%
I}^{\overset{.}{b}J}\\
\left(  D_{F}\right)  _{aI}^{\overset{.}{b}^{\prime}J^{\prime}} &  =0=\left(
D_{F}\right)  _{\overset{.}{a}I}^{b^{\prime}J\prime}%
\end{align}
leaving the components $\left(  D_{F}\right)  _{aI}^{\overset{.}{b}J}$,
$\left(  D_{F}\right)  _{aI}^{b^{\prime}J^{\prime}}$ and $\left(
D_{F}\right)  _{\overset{.}{a}I}^{\overset{.}{b}^{\prime}J^{\prime}}$
arbitrary. These restrictions lead to important constraints on the structure
of the connection that appears in the inner fluctuations of the Dirac
operator. In particular the operator $D$ of the full noncommutative space
given by
\begin{equation}
D=D_{M}\otimes1+\gamma_{5}\otimes D_{F}%
\end{equation}
gets modified to
\begin{equation}
D_{A}=D+A_{\left(  1\right)  }+JA_{\left(  1\right)  }J^{-1}+A_{\left(
2\right)  }%
\end{equation}
where
\begin{equation}
A_{\left(  1\right)  }=%
{\displaystyle\sum}
a\left[  D,b\right]  ,\,\qquad A_{2}=%
{\displaystyle\sum}
\widehat{a}\left[  A_{\left(  1\right)  },\widehat{b}\right]  ,\qquad
\widehat{a}=JaJ^{-1}%
\end{equation}

We have shown in \cite{CCS13b} that components of the connection $A$ which are
tensored with the Clifford gamma matrices $\gamma^{\mu}$ are the gauge fields
of the Pati--Salam model with the symmetry of $SU\left(  2\right)  _{R}\times
SU\left(  2\right)  _{L}\times SU\left(  4\right)  .$ On the other hand, the
non-vanishing components of the connection which are tensored with the gamma
matrix $\gamma_{5}$ are given by
\begin{equation}
\left(  A\right)  _{aI}^{\overset{.}{b}J}\equiv\gamma_{5}
\Sigma _{aI}^{\overset{.}{b}J},\qquad\left(  A\right)  _{aI}%
^{b^{\prime}J^{\prime}}=\gamma_{5}H_{aIbJ},\qquad\left(  A\right)
_{\overset{.}{a}I}^{\overset{.}{b}^{\prime}J^{\prime}}\equiv\gamma
_{5}H_{\overset{.}{a}I\overset{.}{b}J}%
\end{equation}
where $H_{aIbJ}=H_{bJaI}$ and $H_{\overset{.}{a}I\overset{.}{b}J}%
=H_{\overset{.}{b}J\overset{.}{a}I}$, which is the most general Higgs
structure possible. These correspond to the representations with respect to
$SU\left(  2\right)  _{R}\times SU\left(  2\right)  _{L}\times SU\left(
4\right)  :$%
\begin{align}
\Sigma_{aI}^{\overset{.}{b}J} &  =\left(  2_{R},2_{L},1\right)  +\left(
2_{R},2_{L},15\right)  \\
H_{aIbJ} &  =\left(  1_{R},1_{L},6\right)  +\left(  1_{R},3_{L},10\right)  \\
H_{\overset{.}{a}I\overset{.}{b}J} &  =\left(  1_{R},1_{L},6\right)  +\left(
3_{R},1_{L},10\right)
\end{align}
We note, however, that the inner fluctuations form a semi-group and if a
component $\left(  D_{F}\right)  _{aI}^{\overset{.}{b}J}$ or $\left(
D_{F}\right)  _{aI}^{b^{\prime}J^{\prime}}$ or $\left(  D_{F}\right)
_{\overset{.}{a}I}^{\overset{.}{b}^{\prime}J^{\prime}}$ vanish, then the
corresponding $A$ field will also vanish. We can distinguish three cases: 1)
Left-right symmetric Pati--Salam model with fundamental Higgs fields
$\Sigma_{aI}^{\overset{.}{b}J},$ $H_{aIbJ}$ and $H_{\overset{.}{a}%
I\overset{.}{b}J}.$ In this model the field $H_{aIbJ}$ should have a zero vev.
2) A Pati--Salam model where the Higgs field $H_{aIbJ}$ that couples to the
left sector is set to zero which is desirable because there is no symmetry
between the left and right sectors at low energies. 3) If one starts with
$\left(  D_{F}\right)  _{aI}^{\overset{.}{b}J}$ or $\left(  D_{F}\right)
_{aI}^{b^{\prime}J^{\prime}}$ or $\left(  D_{F}\right)  _{\overset{.}{a}%
I}^{\overset{.}{b}^{\prime}J^{\prime}}$ whose values are given by those that
were derived for the Standard Model, then the Higgs fields $\Sigma
_{aI}^{\overset{.}{b}J},$ $H_{aIbJ}$ and $H_{\overset{.}{a}I\overset{.}{b}J}$
will become composite and expressible in terms of more fundamental fields
$\Sigma_{I}^{J},$ $\Delta_{\overset{.}{a}J}$ and $\phi_{\overset{.}{a}}^{b}$ .
We refer to this as the composite model. It has the scalar field $\sigma$ discussed in the previous section as a remnant after spontaneous symmetry breaking \cite{CCS13b}. In fact, contrary to some claims in the literature it is possible to perform the potential analysis in this case in unitarity gauge and arrive at the conclusion that the field content contains the scalar field $\sigma$ ({\em cf.} Appendix \ref{app:potential}).

\begin{figure}
  \centering
\includegraphics[scale=.3]{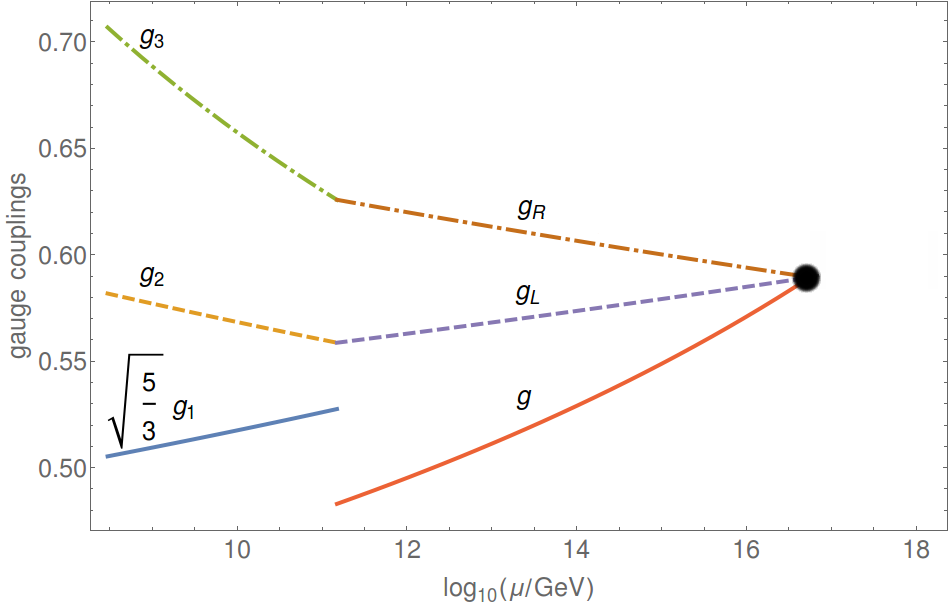}
\caption{Running of the gauge couplings of the Standard Model gauge couplings (below scale $m_R \approx 10^{11} \GeV$) and the Pati--Salam gauge coupling (above scale $m_R$) in case 2.}
\label{fig:ps-running}
  \end{figure}

Depending on the precise particle content we may determine the renormalization group equations of the Pati--Salam gauge couplings $g_{R},g_{L},g$. In \cite{CCS15} we have run them to look for unification of the coupling $g_{R}=g_{L}=g$. The
boundary conditions are taken at the intermediate mass scale $\mu=m_{R}$ to be
the usual (e.g. \cite[Eq. (5.8.3)]{Moh86})
\begin{equation}
\frac{1}{g_{1}^{2}}=\frac{2}{3}\frac{1}{g^{2}}+\frac{1}{g_{R}^{2}},\qquad
\frac{1}{g_{2}^{2}}=\frac{1}{g_{L}^{2}},\qquad\frac{1}{g_{3}^{2}}=\frac
{1}{g^{2}},\label{eq:couplings-relations}%
\end{equation}
in terms of the Standard Model gauge couplings $g_{1},g_{2},g_{3}$. At the
mass scale $m_{R}$ the Pati--Salam symmetry is broken to that of the Standard
Model, and we take it to be the same scale that is present in the see-saw
mechanism. It should thus be of the order $10^{11}-10^{13}$GeV. What we have found in \cite{CCS15} (and this was confirmed by others in \cite{AMST15}) is that in all three cases it is possible to achieve grand unification of the couplings, while connecting to Standard Model physics in the broken, low-energy phase. An example of a running of the gauge coupling is illustrated in Figure \ref{fig:ps-running}.

\subsection{Grand symmetry and twisted spectral triples}
In \cite{DLM14} the next-to-next case\footnote{The case $k=3$ was ruled out by physical considerations \cite{DLM14}.} in the list of irreducible geometries in Equation \eqref{eq:classif} was considered: $k=4$. Thus, one considers
\begin{equation}
A_G = M_4(\bH) \oplus M_8(\C); \qquad H_F := \C^{128}.
  \end{equation}
where $128$ is exactly the number of spinor and internal degrees of freedom combined (including the aforementioned fermion quadruplication). The geometry is then
$$
\left( C^\infty(M,A_G), L^2(M) \otimes H_F, D_M + \gamma_M D_F \right)
$$
where one has to assume that the spinor bundle on $M$ has been trivialized to gather the spinor and internal fermionic degrees of freedom in a single Hilbert space $H_F$.

Note that the above geometry is not a direct product of the continuum with a discrete space. In fact, both the algebra and the Dirac operator $D_M$ contain spinor indices. As a consequence the commutator $[D_M, a]$ can become unbounded, thus challenging one of the basic axioms of spectral triples. Instead, it is possible to guarantee that {\em twisted} commutators are bounded so that this example fits in the general framework of twisted spectral triples developed in \cite{CM08}. In \cite{DM14} the authors identify an inner automorphism $\rho = R (\cdot )R$ of $A_G$ such that
$$
[D,a]_\rho = D a - \rho (a) D 
$$
is bounded.

An interesting question that arises at this point is how to generate inner fluctuations of twisted spectral triples. This was analyzed in full detail from a mathematical viewpoint in \cite{LM16,LM17}. One of the intriguing aspects is the self-adjointness of the Dirac operator under fluctuations (even gauge transformations): for this to be respected one has to impose a compatibility between the twist and the fluctuation.

An alternative route was suggested in \cite{DFLM17}. Namely, one may drop the above condition of self-adjointness and instead look for operators that are Krein-self-adjoint, using the Krein structure on the Hilbert space that is induced by the operator $R$ (defining the twist $\rho$). This will have an intriguing appearance of the Lorentzian structure (given by the Krein inner product) from a purely algebraic and Euclidean starting point. Here we also refer to the nice overview given in \cite{Liz18}.

\subsection{Algebraic constraints on the finite geometry}
An interesting question to consider ---in particular in light of theories that go Beyond the Standard Model--- is whether one can {\em derive} the restricted form of the Dirac operator $D_F$ in \eqref{eq:dirac-sm}. We highlight a few approaches to this question that are present in the literature.

First of all, as mentioned already on page \pageref{page:moduli-dirac}, the form of the $D_F$ in terms of the matrices $Y_\nu, Y_e, Y_u, Y_d$ and $Y_R$ as in Equations \eqref{eq:yukawa-l} and \eqref{eq:yukawa-q} appears naturally in the study of moduli of finite Dirac operators. The only constraint (in addition to the usual conditions layed out in Section \ref{sect:st}) there was that the photon remained massless.

An attempt was made to make the latter condition less {\em ad hoc} is \cite{BF13,BF14,BF18}. They proposed to generalize noncommutative geometry to non-associative noncommutative geometry, thus allowing for non-associative algebras. The crucial idea ---which goes back to Eilenberg--- is to combine the (differential) algebra and (Hilbert space) bimodule into a single algebra, and understand the conditions such as commutant property and first-order conditions as consequences of associativity of the pertinent algebra $B$. However, this associativity is a strong constraint and accordingly further restrict the geometry described by $D_F$. Note that non-associative algebras have also been used in the context of noncommutative geometry and particle physics to predict the number of families (to be three) \cite{TD18}

Another approach to analyzing the form of the Dirac operator $D_F$ by imposing algebraic conditions is taken by \cite{Dab17,DAS18}. Here the authors adopt the principle that, similar to differential forms in the continuum, the finite Hilbert space should be a Morita equivalence between $A$ and the Clifford algebra generated by $A_F$ and $D_F$. One finds that the aforementioned form of $D_F$ does not satisfy this condition but additional entries in $D_F$ should be non-zero. This gives rise to a model Beyond the Standard Model: an analysis of the phenomenological consequences is performed in \cite{KL18,DS18}. In \cite{Ayd19} it was then found that this model does not exhibit grand unification of the Standard Model couplings.

\section{Volume quantization and uniqueness of SM}

In the classification of finite noncommutative spaces we arrived at the result
that the algebra $\mathcal{A}_{F}=\left(  \mathbb{H}_{R}\mathbb{\oplus H}%
_{L}\right)  \oplus M_{4}\left(  \mathbb{C}\right)  $ was the first
possibility out of many of the form $\mathcal{A}_{F}=\left(  M_{n}\left(
\mathbb{H}\right)  _{R}\mathbb{\oplus}M_{n}\left(  \mathbb{H}\right)
_{L}\right)  \oplus M_{4n}\left(  \mathbb{C}\right)  $. in addition we made an
assumption, that seemed arbitrary, of the existence of antilinear isometry
that reduced the algebra $M_{4n}\left(  \mathbb{C}\right)  $ to $\left(
M_{n}\left(  \mathbb{H}\right)  _{R}\mathbb{\oplus}M_{n}\left(  \mathbb{H}%
\right)  _{L}\right)  $. It is necessary to have a stronger evidence of
the uniqueness of our conclusions that helps us to avoid making the above mentioned
assumptions. Surprisingly, the new evidence came in the process of solving a
seemingly completely independent problem, encoding low dimensional geometries,
and in particular dimension four.

\subsection{Higher form of Heisenberg's commutation relations}
Starting with the simple example of one dimensional geometries, consider the
equation
\begin{equation*}
U^{\ast}\left[  D,U\right]  =1,\qquad U^{\ast}U=1
\end{equation*}
where $D$ is self-adjoint operator. Assuming that the one dimensional space is
a closed curve parameterized by coordinate $x$ and the Dirac operator to be
$D=-i\frac{d}{dx}+\alpha$ the above equation simplifies to
\begin{equation*}
-iU^{\ast}dU=dx
\end{equation*}
Writing $U=e^{in\theta}$ we obtain $dx=nd\theta.$ Integrating both sides
implies that the length of the one dimensional curve is an integer multiple of
$2\pi$, the length of $S^1$%
\begin{equation*}%
{\displaystyle\oint\limits_{C}}
dx=n\left(  2\pi\right)
\end{equation*}
To adopt this construction to higher dimensions, we note that we can
characterize the circle $S^{1}$ by the equation $Y^{A}Y^{A}=1,$ $A=1,2$,
$Y^{A\ast}=Y^{A}.$ Assembling the two coordinates $Y^{1},$ $Y^{2}$ in one
matrix, define $Y=Y^{A}\Gamma_{A},$ where $\Gamma_{A},$ $A=1,2$ are taken to
be $2\times2.$ In addition we identify $\Gamma_{1}=\sigma_{1},$ $\Gamma
_{2}=\sigma_{2},$ the Pauli matrices, and define $\Gamma=-i\Gamma_{1}%
\Gamma_{2}=\sigma_{3}$ so that $\Gamma_{+}=\frac{1}{2}\left(  1+\Gamma\right)
$ is a projection operator. We notice that we can write
\[
Y=\left(
\begin{array}
[c]{cc}%
0 & Y^{1}-iY^{2}\\
Y^{1}+iY^{2} & 0
\end{array}
\right)  =\left(
\begin{array}
[c]{cc}%
0 & U^{\ast}\\
U & 0
\end{array}
\right)
\]
where $U=Y^{1}-iY^{2}$ and $U^{\ast}U=1.$ The expression
\begin{equation*}
\left\langle \Gamma_{+}Y\left[  D,Y\right]  \right\rangle =1\label{oned}%
\end{equation*}
where $\left\langle {}\right\rangle $ is defined to be the trace over the
Clifford algebra defined by $\Gamma_{A},$ gives back the equation $U^{\ast
}\left[  D,U\right]  =1.$

For higher dimensional geometries we consider a Riemannian manifold with
dimension $n$ and where the algebra $\mathcal{A}$ is taken to be $C^{\infty
}\left(  M\right)  ,$ the algebra of continuously differentiable functions,
while the operator $D$ is identified with the Dirac operator given by
\begin{equation*}
D_{M}=\gamma^{\mu}\left(  \frac{\partial}{\partial x^{\mu}}+\omega_{\mu
}\right)  ,
\end{equation*}
where $\gamma^{\mu}=e_{a}^{\mu}\gamma^{a}$ and $\omega_{\mu}=\frac{1}{4}%
\omega_{\mu bc}\gamma^{bc}$ is the $SO(n)$ Lie-algebra valued spin-connection
with the (inverse) vielbein $e_{a}^{\mu}$ being the square root of the
(inverse) metric $g^{\mu\nu}=e_{a}^{\mu}\delta^{ab}e_{b}^{\nu}.$ The gamma
matrices $\gamma^{a}$ are anti-hermitian $\left(  \gamma^{a}\right)  ^{\ast
}=-\gamma^{a}$ that define the Clifford algebra $\left\{  \gamma^{a},\gamma
^{b}\right\}  =-2\delta^{ab}.$ The Hilbert space $\mathcal{H}$ is the space of
square integrable spinors $L^{2}\left(  M,S\right)  .$ The chirality operator
$\gamma$ in even dimensions is then given by
\begin{equation*}
\gamma=\left(  i\right)  ^{\frac{n}{2}}\gamma^{1}\gamma^{2}\cdots\gamma^{n}%
\end{equation*}
Starting with manifolds of dimension $2$ we first define the two sphere by the
equation $Y^{A}Y^{A}=1,$ $A=1,2,3$, $Y^{A\ast}=Y^{A}.$ Assembling the three
coordinates $Y^{1},$ $Y^{2},$ $Y^{3}$ in one matrix, defining $Y=Y^{A}%
\Gamma_{A},$ where $\Gamma_{A},$ $A=1,2,3$ are taken to be $2\times2$ Pauli
matrices. Notice that in this case $\Gamma\equiv-i\Gamma_{1}\Gamma_{2}%
\Gamma_{3}=1$ and to generalize equation (\ref{oned}) to two dimensions the
factor $\Gamma$ can be dropped, and we write instead%
\begin{equation*}
\frac{1}{2!}\left\langle Y\left[  D,Y\right]  ^{2}\right\rangle =\gamma
\label{heisenberg2}%
\end{equation*}
The reason we have to include the chirality operator $\gamma$ on the two
dimensional manifold $M$ is that the Dirac operator $D$ appears twice yielding
a product of the form $\gamma_{1}\gamma_{2}=-i\gamma.$ A simple calculation
shows that the above equation in component form is given by
\begin{equation*}
\frac{1}{2!}\epsilon^{\mu\nu}\epsilon_{ABC}Y^{A}\partial_{\mu}Y^{B}%
\partial_{\nu}Y^{C}=\det\left(  e_{\mu}^{a}\right)
\end{equation*}
which is a constraint on the volume form of $M_{2}.$ This implies that the
volume of $M_{2}$ will be an integer multiple of the area of the unit $2$-sphere%
\begin{align*}%
{\displaystyle\int\limits_{M_{2}}}
d^{2}x\sqrt{g}  & =%
{\displaystyle\int}
\epsilon_{ABC}Y^{A}dY^{B}dY^{C}\\
& =n(4\pi)
\end{align*}
where $n$ is the winding number. An example of a map $Y$ with winding number
$n$ is
\begin{equation*}
Y\equiv Y^{1}+iY^{2}=\frac{2z^{n}}{\left\vert z\right\vert ^{2n}+1},\qquad
Y^{3}=\frac{\left\vert z\right\vert ^{2n}-1}{\left\vert z\right\vert ^{2n}%
+1},\qquad z=x^{1}+ix^{2}%
\end{equation*}
From this we deduce that the pullback $Y^{\ast}\left(  w_{n}\right)  $ is a
differential form that does not vanish anywhere. This in turn implies that the
Jacobian of the map $Y$ does not vanish anywhere, and that $Y$ is a covering
of the sphere. The sphere is simply connected, and on each connected component
$M_{j}\subset M_{n}$, the restriction of the map $Y$ to $M_{j}$ is a
diffeomorphism, implying that the manifold must be disconnected, with each
piece having the topology of a sphere. To allow for two dimensional manifolds
with arbitrary topology, our first observation is that condition
(\ref{heisenberg2}) involves the commutator of the Dirac operator $D$ and the
coordinates $Y.$ In momentum space $D$ is the Feynman-slashed $\gamma^{\mu
}p_{\mu}$ momentum and $Y$ are the Feynman-slashed coordinates. This suggests
that the quantization condition is a higher form of Heisenberg commutation
relation quantizing the phase space formed by coordinates and momenta. We
first notice that although the quantization condition is given in terms of the
noncommutative data, the operator $J$ is the only one missing. We therefore
modify the condition to take $J$ into account. The operator $J$ transforms $Y$
into its commutant $Y^{\prime}=iJYJ^{-1}$ so that $\left[  Y,Y^{\prime
}\right]  =0$. Thus let $Y=Y^{A}\Gamma_{A}$ and $Y^{\prime}=iJYJ^{-1}$ and
$\Gamma_{A}^{\prime}=iJ\Gamma_{A}J^{-1}$ so that we can write
\begin{equation*}
Y=Y^{A}\Gamma_{A},\qquad Y^{\prime}=Y^{\prime A}\Gamma_{A}^{\prime},
\end{equation*}
satisfying $Y^{2}=1$ and $Y^{\prime2}=1$ with the Clifford algebras $C_{\pm}$%
\begin{align}
\left\{  \Gamma_{A},\Gamma_{B}\right\}   &  =2\,\delta_{AB},\ \quad(\Gamma
_{A})^{\ast}=\Gamma_{A}\label{Cplus}\\
\left\{  \Gamma_{A}^{\prime},\Gamma_{B}^{\prime}\right\}   &  =-2\,\delta
_{AB},\ \quad(\Gamma_{A}^{\prime})^{\ast}=-\Gamma_{A}^{\prime}\label{Cminus}%
\end{align}
We immediately see that the Clifford algebra $C_{+}=M_{2}\left(
\mathbb{C}\right)  $ and $C_{-}=\mathbb{H}.$ We then define the projection
operator $e=\frac{1}{2}\left(  1+Y\right)  $ satisfying $e^{2}=e$ and
similarly $e^{\prime}=\frac{1}{2}\left(  1+Y^{\prime}\right)  $ satisfying
$e^{\prime2}=e^{\prime}.$ From the tensor product of $E=ee^{\prime}$
satisfying $E^{2}=E,$ we construct $Z=2E-1$ satisfying $Z^{2}=1$ and allowing
us to write
\begin{equation*}
\frac{1}{2}\left\langle Z\left[  D,Z\right]  ^{2}\ \right\rangle =\gamma
\end{equation*}
A straightforward calculation reveals that this relation splits as the sum of
two non-interfering parts%
\begin{equation*}
\frac{1}{2}\left\langle Y\left[  D,Y\right]  ^{2}\right\rangle +\frac{1}%
{2}\left\langle Y^{\prime}\left[  D,Y^{\prime}\right]  ^{2}\right\rangle
=\gamma
\end{equation*}
which in component form reads%
\begin{equation*}
\frac{1}{2!}\epsilon^{\mu\nu}\epsilon_{ABC}\left(  Y^{A}\partial_{\mu}%
Y^{B}\partial_{\nu}Y^{C}+Y^{^{\prime}A}\partial_{\mu}Y^{^{\prime}B}%
\partial_{\nu}Y^{^{\prime}C}\right)  =\det\left(  e_{\mu}^{a}\right)
\end{equation*}
We will show later, when considering the four dimensional case that this
modification allows to reconstruct two dimensional manifolds of arbitrary
topology from the pullbacks of the maps $Y,$ $Y$'.

For three dimensional manifolds $\gamma=1$ and in analogy with the
one-dimensional case we write
\begin{equation*}
\frac{1}{3!}\left\langle \Gamma_{+}Y\left[  D,Y\right]  ^{3}\right\rangle
=1\label{universal}%
\end{equation*}
where $Y=Y^{A}\Gamma_{A},$ $A=1,\ldots4,$ $Y^{2}=1,$ $Y=Y^{\ast},$ $\Gamma
_{A}$ are $4\times4$ Clifford algebra matrices $C_{+}$ where $\left\{
\Gamma_{A},\Gamma_{B}\right\}  =2\,\delta_{AB}$. In this representation of the
$\Gamma$ matrices we have $\Gamma=\Gamma_{5}=\Gamma_{1}\Gamma_{2}\Gamma
_{3}\Gamma_{4}=\left(
\begin{array}
[c]{cc}%
1_{2} & 0\\
0 & -1_{2}%
\end{array}
\right)  $ so that $\Gamma_{+}=\frac{1}{2}\left(  1+\Gamma\right)  $ is a
projection operator. In $d=3,$ we can write
\[
Y=Y^{A}\Gamma_{A}=\left(
\begin{array}
[c]{cc}%
0 & U^{\ast}\\
U & 0
\end{array}
\right)
\]
where $U$ is a unitary $2\times2$ matrix such that it could be written in the
form $U=\exp\left(  i\left(  \alpha_{0}1+\alpha_{a}\sigma^{a}\right)  \right)
$ so that $U^{\ast}U=1$. It is easy to check that $\left\langle Y\left[
D,Y\right]  ^{3}\right\rangle =0$ and that the component form of the above
relation is
\[
\det\left(  e_{\mu}^{a}\right)  =\frac{1}{3!}\epsilon^{\mu\nu\rho}%
\mathrm{Tr}\left(  U^{\ast}\partial_{\mu}UU^{\ast}\partial_{\nu}UU^{\ast
}\partial_{\rho}U\right)
\]
whose integral is the winding number of the $SU(2)$ group manifold. Again,
using the reality operator $J$ we act on the Clifford algebra $Y^{\prime}=iJYJ^{-1}$ so that $\left[  Y,Y^{\prime}\right]  =0$, then $\Gamma_{A}^{\prime}=iJ\Gamma_{A}J^{-1}$ satisfies $\left\{  \Gamma_{A}^{\prime
},\Gamma_{B}^{\prime}\right\}  =-2\,\delta_{AB},$ $(\Gamma_{A}^{\prime}%
)^{\ast}=-\Gamma_{A}^{\prime}$. Forming the projection operators $e=\frac
{1}{2}\left(  1+Y\right)  ,$ $e^{\prime}=\frac{1}{2}\left(  1+Y^{\prime
}\right)  $, we form the tensor product $E=ee^{\prime}$ we define the field
$Z=2E-1,$ and thus the two sided relation becomes
\begin{equation*}
\frac{1}{3!}\left\langle \Gamma_{+}\Gamma_{+}^{\prime}Z\left[  D,Z\right]
^{3}\right\rangle =1
\end{equation*}
A lengthy calculation shows that the component form of this relation separates
into two parts without interference terms%
\begin{align*}
\det\left (  e_{\mu}^{a}\right) & =\frac{1}{3!}\epsilon^{\mu\nu\rho}\bigg(
\mathrm{Tr}\left(  U^{\ast}\partial_{\mu}UU^{\ast}\partial_{\nu}UU^{\ast
}\partial_{\rho}U\right)\\
&\qquad  \qquad +\mathrm{Tr}\left(  U^{^{\prime}\ast}\partial_{\mu
}U^{\prime}U^{^{\prime}\ast}\partial_{\nu}U^{\prime}U^{^{\prime}\ast}%
\partial_{\rho}U^{\prime}\right)  \bigg)
\end{align*}
Finally, for four dimensional manifolds the Clifford algebras
$C_{+}$ and $C_{-}$ defined as in (\ref{Cplus}) (\ref{Cminus}) with
$\Gamma_{A},$ $\Gamma_{A}^{\prime}$, $A=1,\cdots,5$ are known to be given by
$C_{+}=M_{2}\left(  \mathbb{H}\right)  $ and $C_{-}=M_{4}\left(
\mathbb{C}\right)  .$ The quantization condition takes the same form as the
two dimensional case%
\begin{equation*}
\frac{1}{4!}\left\langle Z\left[  D,Z\right]  ^{4}\ \right\rangle
=\gamma\label{Heisenberg}%
\end{equation*}
This relation separates into two non-interfering terms
\begin{equation*}
\frac{1}{4!}\left\langle Y\left[  D,Y\right]  ^{4}\ \right\rangle +\frac
{1}{4!}\left\langle Y^{\prime}\left[  D,Y^{\prime}\right]  ^{4}\ \right\rangle
=\gamma
\end{equation*}
the component form of which is given by
\begin{align*}
\det\left(  e_{\mu}^{a}\right)  &=\frac{1}{4!}\epsilon^{\mu\nu\kappa\lambda
}\epsilon_{ABCDE}\bigg(  Y^{A}\partial_{\mu}Y^{B}\partial_{\nu}Y^{C}%
\partial_{\kappa}Y^{D}\partial_{\lambda}Y^{E}\\
&\qquad \qquad \qquad +Y^{^{\prime}A}\partial_{\mu
}Y^{^{\prime}B}\partial_{\nu}Y^{^{\prime}C}\partial_{\kappa}Y^{^{\prime}%
D}\partial_{\lambda}Y^{^{\prime}E}\bigg)
\end{align*}
One can verify that similar considerations fail when the dimension of the
manifold $n>4$ as there are interference terms between the $Y$ and $Y^{\prime
}.$ Integrating both sides imply%
\begin{equation*}%
{\displaystyle\int\limits_{M_{4}}}
d^{4}x\sqrt{g}=\frac{8}{3}\pi^{2}\left(  N+N^{\prime}\right)
\end{equation*}
where $N$, $N^{\prime}$ are the winding numbers of the two maps $Y,$
$Y^{\prime}.$ An example of a map $Y$ with winding number $n$ is given by
\begin{align*}
Y  & \equiv Y^{4}1+Y^{i}e_{i}=\frac{2x^{n}}{x^{n}\overline{x}^{n}+1},\\
Y^{5}  & =\frac{x^{n}\overline{x}^{n}-1}{x^{n}\overline{x}^{n}+1},
\end{align*}
where $x=x^{4}1+x^{i}e_{i}$ and $e_{i},$ $i=1,2,3$ are the quaternionic
complex structures $e_{i}^{2}=-1,$ $e_{i}e_{j}=\epsilon_{ijk}e_{k},$ $i\neq
j.$

\subsection{Volume quantization}
Consider the smooth maps $\phi_{\pm}:M_{n}\rightarrow S^{n}$ then their
pullbacks $\phi_{\pm}^{\ast}$ would satisfy
\begin{equation*}
\phi_{+}^{\ast}\left(  \alpha\right)  +\phi_{-}^{\ast}\left(  \alpha\right)
=\omega, \label{integer}%
\end{equation*}
where $\alpha$ is the volume form on the unit sphere $S^{n}$ and
$\omega\left(  x\right)  $ is an $n-$form that does not vanish anywhere on
$M_{n}.$ We have shown that for a compact connected smooth oriented manifold
with $n<4$ one can find two maps $\phi_{+}^{\ast}\left(  \alpha\right)  $ and
$\phi_{-}^{\ast}\left(  \alpha\right)  $ whose sum does not vanish anywhere,
satisfying equation (\ref{integer}) such that $%
\int \omega\in\mathbb{Z}.$ The proof for $n=4$ is more difficult and there is an
obstruction unless the second Stieffel--Whitney class $w_{2}$ vanishes, which
is satisfied if $M$ is required to be a spin-manifold and the volume to be
larger than or equal to five units. The key idea in the proof is to note that
the kernel of the Jacobian of the map $Y$ is a hypersurface $\Sigma$ of co-dimension $2$ and
therefore
\begin{equation*}
\dim\Sigma=n-2.
\end{equation*}
We can then construct a map $Y^{\prime}=Y\circ\psi$ where $\psi$ is a
diffeomorphism on $M$ such that the sum of the pullbacks of $Y$ and
$Y^{\prime}$ does not vanish anywhere. \ The coordinates $Y$ are defined over
a Clifford algebra $C_{+}$ spanned by $\left\{  \Gamma_{A},\Gamma_{B}\right\}
=2\delta_{AB}.$ For $n=2$, $C_{+}=M_{2}\left(  \mathbb{C}\right)  $ while for
$n=4$, $C_{+}=M_{2}\left(  \mathbb{H}\right)  \oplus M_{2}\left(
\mathbb{H}\right)  $ where $\mathbb{H}$ is the field of quaternions. However,
for $n=4,$ since we will be dealing with irreducible representations we take
$C_{+}=M_{2}\left(  \mathbb{H}\right)  .$ Similarly the coordinates
$Y^{\prime}$ are defined over the Clifford algebra $C_{-}$ spanned by
$\left\{  \Gamma_{A}^{\prime},\Gamma_{B}^{\prime}\right\}  =-2\delta_{AB}$ and
for $n=2$, $C_{-}=\mathbb{H\oplus H}$ and for $n=4$, $C_{-}=M_{4}\left(
\mathbb{C}\right)  .$ The operator $J$ acts on the two algebras $C_{+}\oplus
C_{-}$ in the form $J\left(  x,y\right)  =\left(  y^{\ast},x^{\ast}\right)  $
(i.e. it exchanges the two algebras and takes the Hermitian conjugate). The
coordinates $Z=\frac{1}{2}\left(  Y+1\right)  \left(  Y^{\prime}+1\right)
-1,$ then define the matrix algebras \cite{CCM14}
\begin{align*}
\mathcal{A}_{F}  &  =M_{2}\left(  \mathbb{C}\right)  \oplus\mathbb{H},\qquad
n=2\\
\mathcal{A}_{F}  &  =M_{2}\left(  \mathbb{H}\right)  \oplus M_{4}\left(
\mathbb{C}\right)  ,\qquad n=4.
\end{align*}
One, however, must remember that the maps $Y$ and $Y^{\prime}$ are functions
of the coordinates of the manifold $M$ and therefore the algebra associated
with this space must be
\begin{align*}
\mathcal{A}  &  =C^{\infty}\left(  M,\mathcal{A}_{F}\right) \\
&  =C^{\infty}\left(  M\right)  \otimes\mathcal{A}_{F}.
\end{align*}
To see this consider, for simplicity, the $n=2$ case with only the map $Y.$
The Clifford algebra $C_{-}=\mathbb{H}$ is spanned by the set $\left\{
1,\Gamma^{A}\right\}  ,$ $A=1,2,3,$ where $\left\{  \Gamma^{A},\Gamma
^{B}\right\}  =-2\delta^{AB}.$ We then consider functions which are made out
of words of the variable $Y$ formed with the use of constant elements of the
algebra \cite{C00}
\[%
{\displaystyle\sum\limits_{i=1}^{\infty}}
a_{1}Ya_{2}Y\cdots a_{i}Y,\qquad a_{i}\in\mathbb{H},
\]
which will generate arbitrary functions over the manifold which is the most
general form since $Y^{2}=1$. One can easily see that these combinations
generate all the spherical harmonics. This result could be easily generalized
by considering functions of the fields
\[
Z=\frac{1}{2}\left(  Y+1\right)  \left(  Y^{\prime}+1\right)  -1,\qquad
Y\in\mathbb{H},\quad Y^{\prime}\in M_{2}\left(  \mathbb{C}\right)  ,
\]
showing that the noncommutative algebra generated by the constant matrices and
the Feynman slash coordinates $Z$ is given by \cite{C00}
\[
\mathcal{A}=C^{\infty}\left(  M_{2}\right)  \otimes\left(  \mathbb{H+}%
M_{2}\left(  \mathbb{C}\right)  \right)  .
\]
We now restrict ourselves to the physical case of $n=4.$ Here the algebra is
given by
\begin{equation*}
\mathcal{A}=C^{\infty}\left(  M_{4}\right)  \otimes\left(  M_{2}%
(\mathbb{H})\mathbb{+}M_{4}\left(  \mathbb{C}\right)  \right)  .
\end{equation*}
The associated Hilbert space is
\begin{equation*}
\mathcal{H}=L^{2}\left(  M_{4},S\right)  \otimes\mathcal{H}_{F}.
\end{equation*}
The Dirac operator mixes the finite space and the continuous manifold
non-trivially%
\begin{equation*}
D=D_{M}\otimes1+\gamma_{5}\otimes D_{F},
\end{equation*}
where $D_{F\text{ }}$ is a self adjoint operator in the finite space. The
chirality operator is
\begin{equation*}
\gamma=\gamma_{5}\otimes\gamma_{F},
\end{equation*}
and the anti-unitary operator $J$ is given by
\begin{equation*}
J=J_{M}\gamma_{5}\otimes J_{F},
\end{equation*}
where $J_{M}$ is the charge-conjugation operator $C$ on $M$ and $J_{F}$ the
anti-unitary operator for the finite space. Thus an element $\Psi
\in\mathcal{H}$ is of the form $\Psi=\left(
\begin{array}
[c]{c}%
\psi_{A}\\
\psi_{A^{\prime}}%
\end{array}
\right)  $ where $\psi_{A}$ is a $16$ component $L^{2}\left(  M,S\right)  $
spinor in the fundamental representation of $\mathcal{A}_{F}$ of the form
$\psi_{A}=\psi_{\alpha I}$ where $\alpha=1,\cdots,4$ with respect to
$M_{2}\left(  \mathbb{H}\right)  $ and $I=1,\cdots,4$ with respect to
$M_{4}\left(  \mathbb{C}\right)  $ and where $\psi_{A^{\prime}}=C\psi
_{A}^{\ast}$ is the charge conjugate spinor to $\psi_{A}$ \cite{CC10}. The
chirality operator $\gamma$ must commute with elements of $\mathcal{A}$ which
implies that $\gamma_{F}$ must commute with elements in $\mathcal{A}_{F}.$
Commutativity of the chirality operator $\gamma_{F}$ with the algebra
$\mathcal{A}_{F}$ and that this $\mathbb{Z}/2$ grading acts non-trivially
reduces the algebra $M_{2}\left(  \mathbb{H}\right)  $ to $\mathbb{H}%
_{R}\oplus\mathbb{H}_{L}$ \cite{CCM14}. Thus the $\gamma_{F}$ is identified with
$\gamma_{F}=\Gamma^{5}=\Gamma^{1}\Gamma^{2}\Gamma^{3}\Gamma^{4}$ and the
finite space algebra reduces to
\begin{equation*}
\mathcal{A}_{F}=\mathbb{H}_{R}\oplus\mathbb{H}_{L}\oplus M_{4}\left(
\mathbb{C}\right)  .
\end{equation*}
This can be easily seen by noting that an element of $M_{2}\left(
\mathbb{H}\right)  $ takes the form $\left(
\begin{array}
[c]{cc}%
q_{1} & q_{2}\\
q_{3} & q_{4}%
\end{array}
\right)  $ where each $q_{i},$ $i=1,\cdots,4,$ is a $2\times2$ matrix
representing a quaternion. Taking the representation of $\Gamma^{5}=\left(
\begin{array}
[c]{cc}%
1_{2} & 0\\
0 & -1_{2}%
\end{array}
\right)  $ to commute with $M_{2}\left(  \mathbb{H}\right)  $ implies that
$q_{2}=0=q_{3},$ thus reducing the algebra to $\mathbb{H}_{R}\oplus
\mathbb{H}_{L}.$ Therefore the index $\alpha=1,\cdots,4$ splits into two
parts, $\overset{.}{a}=\overset{.}{1},\overset{.}{2}$ which is a doublet under
$\mathbb{H}_{R}$ and $a=1,2$ which is a doublet under $\mathbb{H}_{L}$. The
spinor $\Psi$ further satisfies the chirality condition $\gamma\Psi=\Psi$
which implies that the spinors $\psi_{\overset{.}{a}I}$ are in the $\left(
2_{R},1_{L},4\right)  $ with respect to the algebra $\mathbb{H}_{R}%
\mathbb{\oplus H}_{L}\oplus M_{4}\left(  \mathbb{C}\right)  $ while $\psi
_{aI}$ are in the $\left(  1_{R},2_{L},4\right)  $ representation. 
The finite space Dirac operator $D_{F}$ is then a $32\times32$ Hermitian matrix
acting on the $32$ component spinors $\Psi.$ In addition we take three copies
of each spinor to account for the three families, but will omit writing an
index for the families. At present we have no explanation for why the number
of generations should be three. The Dirac operator for the finite space is
then a $96\times96$ Hermitian matrix. The Dirac action is then given by
\cite{CCM07}
\begin{equation*}
\left(  J\Psi,D\Psi\right)  .
\end{equation*}
We note that we are considering compact spaces with Euclidean signature and
thus the condition $J\Psi=\Psi$ could not be imposed. It could, however, be
imposed if the four dimensional space is Lorentzian \cite{Bar06}.The reason
is that the $KO$ dimension of the finite space is $6$ because the operators
$D_{F},$ $\gamma_{F}$ and $J_{F}$ satisfy%
\begin{equation*}
J_{F}^{2}=1,\qquad J_{F}D_{F}=D_{F}J_{F},\qquad J_{F}\gamma_{F}=-\gamma
_{F}J_{F}.
\end{equation*}
The operators $D_{M},$ $\gamma_{M}=\gamma_{5},$ and $J_{M}=C$ for a compact
manifold of dimension $4$ satisfy
\begin{equation*}
J_{M}^{2}=-1,\qquad J_{M}D_{M}=D_{M}J_{M},\qquad J_{M}\gamma_{5}=\gamma
_{5}J_{M}. \label{Euclidean}%
\end{equation*}
Thus the $KO$ dimension of the full noncommutative space $\left(
\mathcal{A},\mathcal{H},D\right)  $ with the decorations $J$ and $\gamma$
included is $10$ and satisfies
\begin{equation*}
J^{2}=-1,\qquad JD=DJ,\qquad J\gamma=-\gamma J.
\end{equation*}
We have shown in \cite{CCM07} that the path integral of the Dirac action,
thanks to the relations $J^{2}=-1$ and $J\gamma=-\gamma J$, yields a
Pfaffian of the operator $D$ instead of its determinant and thus eliminates
half the degrees of freedom of $\Psi$ and have the same effect as imposing the
condition $J\Psi=\Psi.$

We have also seen that the operator $J$ sends the algebra $\mathcal{A}$ to its
commutant, and thus the full algebra acting on the Hilbert space $\mathcal{H}$
is $\mathcal{A\otimes A}^{o}.$ Under automorphisms of the algebra
\begin{equation*}
\Psi\rightarrow U\Psi,
\end{equation*}
where $U=u\widehat{u}$ with $u\in\mathcal{A},$ $\widehat{u}\in\mathcal{A}^{o}$
with $\left[  u,\widehat{u}\right]  =0$, it is clear that Dirac action is not invariant.

At this point it is clear that we have retrieved all our conclusions we have
before arriving at a unique possibility, which is to have a noncommutative
space corresponding to the Pati--Salam Model we considered before, and in the special
case where the Dirac operator and algebra satisfy the order one condition, the
result is the noncommutative space of the Standard Model. We have thus
succeeded in obtaining the Pati--Salam Model and Standard Model as unique
possibilities starting with the two sided Heisenberg like equation
(\ref{Heisenberg}) thus eliminating all other possibilities obtained in
classifying finite noncommutative spaces of KO dimension $6.$ There is no need
to assume the existence of an isometry that reduces the first algebra from
$M_{4}\left(  \mathbb{C}\right)  $ to $M_{2}\left(  \mathbb{H}\right)  $, and
no need to assume that the KO dimension of the finite space to be $6.$ These
results are very satisfactory and serve to enhance our confidence of the fine
structure of space time as given by the above derived noncommutative space.

\section{Outlook: towards quantization}
   
Starting with the simple observation that the Higgs field could be interpreted
as the link between two parallel sheets separated by a distance of the order
of $10^{-16}$ cm  it took enormous effort to identify a noncommutative space
where the spectrum of the Standard Model could fit. Small deviations from the
model, such as the need for a real structure and a KO dimension $6$, were
taken as input to fine tune and determine precisely the noncommutative space.
The spectral action principle proved to be very efficient way in evaluating
the bosonic sector of the theory. Having identified the noncommutative space,
the next target was to understand why nature would chose the Standard Model
and not any other possibility. A classification of finite spaces revealed the
special nature of the the finite part of the noncommutative space identified.
Work on encoding manifolds with dimensions equal to four satisfying a higher
form of Heisenberg type equation showed that the most general solution of this
equation is that of a noncommutative space which is a product of a
four-dimensional Riemannian spin-manifold times the finite space corresponding
to a Pati--Salam unification model. The Standard Model is a special case of
this space where a first order differential condition is satisfied. After a
long journey the reasons why nature chose the Standard Model is  now reduced
to determining solutions of a higher form of Heisenberg equation. With such
little input, it is quite satisfying to learn that it is possible to answer
many of the questions which puzzled theorists for a long time. We now know why
there are 16 fermions per generation, why the gauge group is $SU\left(
3\right)  \times SU\left(  2\right)  \times U\left(  1\right)  ,$ an
explanation of the Higgs field and origin of spontaneous symmetry breaking.
The Spectral model also predicts a Majorana mass for the right-handed
neutrinos and explains the see-saw mechanism. We thus understand unification
of all fundamental forces as a geometrical theory based on the spectral action
principle of a noncommutative space. 

Naturally, there are many questions that are still unanswered, and this
motivates the need for further research to address these problems using
noncommutative geometry considerations. To conclude, we mention few of the
possible directions of future research. One important aspect to consider is
the renormalizability properties of the spectral model. Another problem is to
study the quantum properties of the Dirac operator and whether it could be
related to the pullbacks of the maps used in determining the quanta of
geometry. The future of noncommutative geometry in the program of unification
of all fundamental interactions looks now to be very promising.

\appendix

\section{Pati--Salam model: potential analysis}
\label{app:potential}
We here include the scalar potential analysis for the composite Pati--Salam model, as described in Section \ref{sect:patisalam} above. 

If there is unification of lepton and quark couplings, then $\rho=1$ so that the $\Sigma^I_J$-field decouples. In that case we have
\begin{multline*}
\L_{pot} (\phi_{\overset{.}{a}}^{b},\Delta_{\overset{.}{a}I})=
-\mu^2 \phi_{\overset{.}{a}}^{c} \phi_{c}^{\overset{.}{a}}
- \nu^2 \left(  \Delta_{\overset{.}{a}%
K}\overline{\Delta}^{\overset{.}{a}K}\right)  ^{2}%
+ \lambda_{\Sigma} 
\phi_{a}^{\overset{.}{c}}
\phi_{\overset{.}{c}}^{b} 
\phi_{b}^{\overset{.}{d}}
\phi_{\overset{.}{d}}^{a} 
\\ 
 + \lambda_H \left(\Delta_{\overset{.}{a}K}\overline{\Delta}^{\overset{.}{a}L}%
\Delta_{\overset{.}{b}L}\overline{\Delta}^{\overset{.}{b}K} \right)^2
+\lambda_{H \Sigma}
\left(  \Delta_{\overset{.}{a}%
J}\overline{\Delta}^{\overset{.}{a}J}\Delta_{\overset{.}{c}I}\overline{\Delta
}^{\overset{.}{d}I}\right) \phi_{b}^{\overset{.}{c}}  \phi_{\overset{.}{d}}^{b}
\end{multline*}
where we have absorbed some constant factors by redefining the couplings $\lambda_H, \lambda_{H \Sigma}$ and $\lambda_\Sigma$.

We choose unitarity gauge for the $\Delta$ and $\phi$-fields, in the following precise sense. 
\begin{lma}
\label{lma:unitary-gauge1}
For each value of the fields $\{ \phi_{\dot a}^b, \Delta_{\dot a I}\}$ there is an element $(u_R,u_L,u) \in SU(2)_R \times SU(2)_L \times SU(4)$ such that 
\begin{align*}
u_R \begin{pmatrix} \phi_{\dot 1}^1 &  \phi_{\dot 1}^2 \\[1mm] \phi_{\dot 2}^1 &  \phi_{\dot 2}^2 
\end{pmatrix} u_L^* & = \begin{pmatrix} h & 0 \\ 0 & \chi \end{pmatrix}
\intertext{and}
u_R \begin{pmatrix} \Delta_{\dot 1 1}& \Delta_{\dot 1 2}& \Delta_{\dot 1 3}& \Delta_{\dot 1 4}\\ \Delta_{\dot 2 1}& \Delta_{\dot 2 2}& \Delta_{\dot 2 3}& \Delta_{\dot 2 4}\end{pmatrix}  u^t &= \begin{pmatrix} 1+\delta_0 & 0& 0 & 0 \\ \delta_1 & \eta_1& 0 &0  \end{pmatrix}
\end{align*}
where $h ,\delta_0, \delta_1, \eta_1$ are real fields and $\chi$ is a complex field.
\end{lma}
\proof
Consider the singular value decomposition of the $2 \times 2$ matrix $(\phi_{\dot a}^b)$:
\begin{equation*}
 \begin{pmatrix} \phi_{\dot 1}^1 &  \phi_{\dot 1}^2 \\[1mm] \phi_{\dot 2}^1 &  \phi_{\dot 2}^2 
\end{pmatrix} = U  \begin{pmatrix} h & 0 \\ 0 & k \end{pmatrix} V^*
\end{equation*}
for unitary $2 \times 2$ matrices $U,V$ and real coefficients $h,k$. If we define
\begin{align*}
u_R &=   \begin{pmatrix} 1 & 0 \\ 0 & \det U\end{pmatrix}U^* \in SU(2)_R\\
u_L &=   \begin{pmatrix} 1 & 0 \\ 0 & \det V\end{pmatrix} V^*\in SU(2)_L
\end{align*}
it follows that
\begin{align*}
u_R \begin{pmatrix} \phi_{\dot 1}^1 &  \phi_{\dot 1}^2 \\[1mm] \phi_{\dot 2}^1 &  \phi_{\dot 2}^2 
\end{pmatrix} u_L^* & = \begin{pmatrix} h & 0 \\ 0 & k \det UV^* \end{pmatrix}
=: \begin{pmatrix} h & 0 \\ 0 & \chi \end{pmatrix}.
\end{align*}
Next, we consider $\Delta_{\dot aI}$ and write
\begin{equation*}
\left(\Delta_{\dot aI}\right) = \begin{pmatrix} u_1^* \\  u_2^* \end{pmatrix},\qquad \text{with } u_a^* = \begin{pmatrix} \Delta_{\dot a 1}& \Delta_{\dot a 2}& \Delta_{\dot a 3}& \Delta_{\dot a 4} \end{pmatrix} 
\end{equation*}
for $a=1,2$. We may suppose that the vectors $u_1,u_2$ are such that their inner product $u_1^* u_2$ is a real number. Indeed, if this is not the case, then multiply $\Delta_{\dot aI}$ by a matrix in $SU(2)_R$ as follows:
\begin{equation*}
\begin{pmatrix} u_1^* \\  u_2^* \end{pmatrix}\mapsto 
\begin{pmatrix} \alpha & 0  \\ 0 & \alpha^* \end{pmatrix}\begin{pmatrix} u_1^* \\  u_2^* \end{pmatrix} = \begin{pmatrix} \alpha u_1^* \\  \alpha^* u_2^* \end{pmatrix}.
\end{equation*}
Now the inner product is $(\alpha^* u_1)^* \alpha u_2 = (\alpha)^2 u_1^* u_2$ and we may choose $\alpha$ so as to cancel the phase of $u_1^* u_2$. Moreover, this transformation respects the above form of $\phi_{\dot a}^b$ after a $SU(2)_L$-transformation of exactly the same form:
\begin{equation*}
\begin{pmatrix} h & 0 \\ 0 & \chi \end{pmatrix} \mapsto 
\begin{pmatrix} \alpha & 0  \\ 0 & \alpha^* \end{pmatrix} \begin{pmatrix} h & 0 \\ 0 & \chi \end{pmatrix} \begin{pmatrix} \alpha & 0  \\ 0 & \alpha^* \end{pmatrix} ^* = \begin{pmatrix} h & 0 \\ 0 & \chi \end{pmatrix} .
\end{equation*}
Thus let us continue with the vectors $u_1,u_2$ satisfying $u_1^* u_2 \in \R$. We apply Gramm-Schmidt orthonormalization to $u_1$ and $u_2$, to arrive at the following orthonormal set of vectors $\{e_1,e_2\}$ in $\C^4$:
\begin{equation*}
e_1 = \frac{u_1 }{\| u_1\|}; \qquad e_2 = \frac{u_2 - \frac{u_1^* u_2}{\|u_1\|} u_1}{\|u_2 - \frac{u_1^* u_2}{\|u_1\|} u_1\|}.
\end{equation*}
We complete this set by choosing two additional orthonormal vectors $e_3$ and $e_4$ and write a unitary $4 \times 4$ matrix:
\begin{equation*}
U = \begin{pmatrix} e_1 & e_2 & e_3 & e_4 \end{pmatrix}
\end{equation*}
The sought-for matrix $u \in SU(4)$ is determined by 
\begin{equation*}
u^t = U \begin{pmatrix} 1_3 & 0 \\ 0 & \det U^* \end{pmatrix}
\end{equation*}
so as to give 
\begin{equation*}
\left(\Delta_{\dot aI}\right) u^t = \begin{pmatrix} u_1^* e_1 & 0 & 0 & 0  \\  u_2^* e_1 & u_2^* e_2 & 0 & 0  \end{pmatrix} =:  \begin{pmatrix} 1+\delta_0 & 0 & 0 & 0  \\ \delta_1 & \eta_1 & 0 & 0  \end{pmatrix}\qedhere
\end{equation*}
\endproof

\begin{rem}
Note that this is compatible with the dimension of the quotient of the space of field values by the group. Indeed, the fields $\phi_{\dot a}^b$ and $\Delta_{\dot a I}$ span a real 24-dimensional space (at each manifold point). The dimension of the orbit space is then $24- \dim P$ with $P$ a principal orbit of the action of $SU(2)_R \times SU(2)_L \times SU(4)$ on the space of field values. This dimension $ \dim P$ is determined by the dimension of the group and of a principal isotropy group. 

First, we see that up to conjugation there is always a $SU(2)$-subgroup of $SU(4)$ leaving $\Delta_{\dot aI}$ invariant: it corresponds to $SU(2)$-transformations in the space orthogonal to the vectors $\Delta_{\dot 1 I}$ and $\Delta_{\dot 2I}$ in $\C^4$. Moreover, one can compute that the isotropy subgroup of the field values
\begin{equation*}
\begin{pmatrix} \phi_{\dot a}^b \end{pmatrix} = \begin{pmatrix}1 & 0 \\ 0 & 0 \end{pmatrix}; \qquad \begin{pmatrix} \Delta_{\dot a I} \end{pmatrix} = \begin{pmatrix}1 & 0& 0 & 0 \\ 1 & 1 &  0 & 0 \end{pmatrix}
\end{equation*}
 is given by $\mathbb Z_2 \times SU(2)$. Hence, the dimension of the principal orbit is $21 - 3 = 18$ so that the orbit space is 6-dimensional. This corresponds to the 4 real fields $h,\delta_0,\delta_1,\eta_1$ and the complex field $\chi$. 
\end{rem}
We allow for the colour $SU(3)$-symmetry not to be broken spontaneously, hence we only choose unitarity gauge in the $SU(2)_R \times SU(2)_L \times U(1)$-representations. That is, we retain the row vector $\Delta_{\dot 2 I}$ for $I=1,\ldots, 4$ as a variable and write
\begin{equation*}
 \begin{pmatrix} \Delta_{\dot a I} \end{pmatrix} = \begin{pmatrix}\sqrt{w}+\delta_0/\sqrt w & 0& 0 & 0 \\ \delta_1/\sqrt{w} & \eta_1/\sqrt{w} &  \eta_2/\sqrt{w} & \eta_3/\sqrt{w} \end{pmatrix}
\end{equation*}
so that $(\eta_i)$ forms a scalar $SU(3)$-triplet field (so-called {\em scalar leptoquarks}). The reason for the rescaling with $\sqrt{w}$ is that it yields the right kinetic terms for $\delta_0,\delta_1$ and $\eta$. Indeed, from the spectral action we then have 
\begin{align*}
\frac{1}{2}\partial_{\mu}H_{\overset{.}{a}%
I\overset{.}{b}J}\partial^{\mu}H^{\overset{.}{a}I\overset{.}{b}J}
&=\frac{1}{2} \partial_\mu \left(\Delta_{\overset{.}{a}J}\Delta_{\overset{.}{b}I}\right) \partial^\mu \left(\Delta^{\overset{.}{a}J}\Delta^{\overset{.}{b}J}\right) 
\\
&\sim \sum_{a=0}^1 \partial_\mu \delta_a \partial^\mu \delta^a + \partial_\mu \eta \partial^\mu \eta^* + \text{ higher order}
\end{align*}
The scalar potential becomes in terms of the fields $h,\chi,\delta_0,\delta_1, \eta_i$:
\begin{align*}
&\L_{pot}(h,\chi, \delta_0,\delta_1,\eta) = - \mu^2 ( h^2 + |\chi|^2) -\nu^2  \left( (w+ \delta_0)^2  + \delta_1^2 + |\eta|^2 \right)^2 /w^2
\\
\nn &
\quad + \lambda_{H\Sigma}  \left( (w+\delta_0)^2 h^2 + (\delta_1^2 + |\eta|^2) |\chi|^2 \right)\left( (w+ \delta_0)^2  + \delta_1^2 + |\eta|^2 \right) /w^2 \\
\nn &
\quad + \lambda_H \left( (w+\delta_0)^4 + 2 (w+ \delta_0)^2 \delta_1^2+ (\delta_1^2 +|\eta|^2)^2 \right)^2/w^4 +  \lambda_\Sigma (h^4+ |\chi|^4) 
\end{align*}
As we are interested in the truncation to the Standard Model, we look for extrema with $\langle \delta_1\rangle =\langle \eta_i\rangle=0$, whilst setting $\langle h \rangle =v, \langle \delta_0 \rangle = 0, \langle \chi \rangle = x$. Note that the symmetry of these vevs is
\begin{multline*}
\left\{ 
\left(\begin{pmatrix} \lambda & 0 \\ 0 & \lambda^* \end{pmatrix},
\begin{pmatrix} \lambda^* & 0 \\ 0 & \lambda \end{pmatrix},
\begin{pmatrix} \lambda^* & 0 \\ 0 & m \end{pmatrix}\right): \lambda \in U(1) , m \in SU(3)  \right\}
\\\subset SU(2)_R \times SU(2)_L \times SU(4)
\end{multline*}
In other words, $SU(2)_R \times SU(2)_L \times SU(4)$ is broken by the above vevs to $U(1) \times SU(3)$.

The first derivative of $V$ vanishes for these vevs precisely if
\begin{align*}
2v (w^2 \lambda_{H \Sigma} +2v^2 \lambda_\Sigma - \mu^2 ) = 0,\\
4 x^3 \lambda_\Sigma - 2x \mu^2 = 0,\\
4w (2w^2 \lambda_H + v^2 \lambda_{H \Sigma} - \nu^2) =0.
\end{align*}
This gives rise to the fine-tuning of $v,w$ as in \cite{CC12}:
\begin{equation*}
w^2 \lambda_{H \Sigma} +2v^2 \lambda_\Sigma - \mu^2 , \qquad 2w^2 \lambda_H + v^2 \lambda_{H \Sigma} - \nu^2
\end{equation*}
choosing $\mu$ and $\nu$ such that the solutions $v,w$ are of the desired orders. Moreover, we find that the vev for $\chi$ either vanishes or is equal to $x=\sqrt{\mu^2/2\lambda_\Sigma}$. Note that this latter vev appears precisely at the entry $k^d h$ (or $k^e h$) of the finite Dirac operator, which we have disregarded by setting $\rho=1$. 

If $\langle \chi \rangle = x =0$ then the Hessian is (derivatives with respect to $h,\chi,\delta_0,\delta_1, \eta$):
\begin{equation*}
\left(
\begin{smallmatrix}
 8 v^2 \lambda_\Sigma  & 0 & 8 v w \lambda_{H \Sigma} & 0 & 0 \\
 0 & -2 w^2\lambda_{H\Sigma} -4 v^2 \lambda \Sigma  & 0 & 0 & 0 \\
 8 v w \lambda_{H\Sigma} & 0 & 32 w^2 \lambda_H & 0 & 0 \\
 0 & 0 & 0 & -2 v^2 \lambda_{H\Sigma}  & 0 \\
 0 & 0 & 0 & 0 & -8 \lambda_H w^2-2 v^2 \lambda_{H\Sigma} w^2  {\bf 1}_3 \\
\end{smallmatrix}
\right)
\end{equation*}
where the ${\bf 1}_3$ is the identity matrix in colour space, corresponding to the $\eta$-field. This Hessian is not positive definite so we disregard the possibility that $\langle \chi \rangle =0$. 

If $x=\sqrt{\mu^2/2\lambda_\Sigma}$ then the Hessian is
\begin{equation*}
\left(\begin{smallmatrix}
8v^2 \lambda_\Sigma & 0 & 8vw \lambda_{H\Sigma} & 0 & 0 \\
0 & 4w^2 \lambda_{H \Sigma}+8v^2 \lambda_\Sigma & 0 & 0 & 0\\ 
8vw \lambda_{H\Sigma}  & 0 & 32 w^2 \lambda_H & 0 & 0 \\
0 & 0 & 0 & w^2 \frac{\lambda_{H \Sigma}^2}{\lambda_\Sigma} & 0 \\
0 & 0 & 0 & 0 & w^2 \frac{ \lambda_{H\Sigma}^2 - 8 \lambda_H \lambda_\Sigma}{\lambda_\Sigma}{\bf 1}_3
\end{smallmatrix}\right)
\end{equation*}
which is positive-definite if
\begin{equation}
\label{eq:positive-mass}
\lambda_{H\Sigma}^2  \geq 8 \lambda_H \lambda_\Sigma.
\end{equation}
Note that this relation may hold only at high-energies. The masses for $\chi$, $\delta_1$ and $\eta$ are then readily found to be:
\begin{align*}
m_\chi^2 &= 4w^2 \lambda_{H \Sigma}+8v^2 \lambda_\Sigma,\\
m_{\delta_1}^2 &= w^2 \frac{\lambda_{H \Sigma}^2}{\lambda_\Sigma} ,\\
m_{\eta}^2 &=w^2 \frac{ \lambda_{H\Sigma}^2 - 8 \lambda_H \lambda_\Sigma}{\lambda_\Sigma} .
\end{align*}
Under the assumption that $v^2 \approx 10^2 \GeV, w^2 \approx 10^{11} \GeV$ we have $m_\chi^2 \approx 10^{11} \GeV$ and $m_{\delta_1}^2, m_{\eta} \approx 10^{11}\GeV$. 

The (non-diagonal) $h$ and $\delta_0$ sector has mass eigenstates as in \cite{CC12}:
\begin{multline*}
m_\pm^2 = 16 w^2\lambda_H +4 v^2 \lambda_\Sigma \\\pm 4\sqrt{16 w^{4} \lambda_H^2 + v^4 \lambda_\Sigma^2 + 4v^2 w^2 \left(\lambda_{H \Sigma}^2 - 2 \lambda_H \lambda_\Sigma \right)}
\end{multline*}
Under the assumption that $v^2 \ll w^2$ we can expand the square root:
\begin{align*}
&4 \sqrt{ 16 \lambda_H^2 w^{4}  \left(1 + \frac{\lambda_\Sigma^2}{\lambda_H^2}\frac{v^4}{w^{4}} +  \frac{\lambda_{H \Sigma}^2 - 2 \lambda_H \lambda_\Sigma}{4\lambda_H^2} \frac{v^2}{w^2}\right)}\\ \nn
&\qquad\approx 16 \lambda_H  w^2 \left(1 +\frac{\lambda_{H \Sigma}^2 - 2 \lambda_H \lambda_\Sigma}{8\lambda_H^2} \frac{v^2}{w^2}\right)
\\\nn
&\qquad=  16\lambda_H  w^2  +\frac{2\lambda_{H \Sigma}^2}{\lambda_H} v^2 
-  4\lambda_\Sigma v^2.
\end{align*}
Consequently, 
\begin{align*}
m_+ &\approx 32  \lambda_H w^2 + 2 \frac{2\lambda_{H \Sigma}^2}{\lambda_H} v^2, \\
m_- &\approx 8 \lambda_\Sigma v^2 \left(1- \frac{\lambda_{H \Sigma}^2}{4\lambda_H \lambda_\Sigma} \right).
\end{align*}
which are of the order of $10^{11}$ and $10^2 \GeV$, respectively. This requires that we have at low energies
\begin{equation}
\label{eq:positive-higgs-mass}
4\lambda_H \lambda_\Sigma \geq \lambda_{H \Sigma}^2,
\end{equation}
which fully agrees with \cite{CC12} when we identify $\delta_0 \equiv \sigma$ and with the couplings related via
\begin{equation*}
\lambda_H = \frac14 \lambda_\sigma , \qquad \lambda_{H \Sigma} = \frac12 \lambda_{h\sigma}, \qquad \lambda_\Sigma = \frac14 \lambda_h
\end{equation*}
Note the tension between Equations \eqref{eq:positive-higgs-mass} and \eqref{eq:positive-mass}, calling for a careful study of the running of the couplings in order to guarantee positive mass eigenstates at their respective energies.

We have summarized the scalar particle content of the above model in Table \ref{table:part-cont}.
\begin{table}
$$
\begin{array}{c||ccc}
& U(1)_Y & SU(2)_L & SU(3) \\
\hline\hline
\begin{pmatrix}\phi_1^0 \\ \phi_1^+ \end{pmatrix} = \begin{pmatrix} \phi_{\dot 1}^1\\ \phi_{\dot 1}^2 \end{pmatrix} & 1 & 2 & 1 \\[4mm]
\begin{pmatrix}\phi_2^- \\ \phi_2^0 \end{pmatrix} = \begin{pmatrix} \phi_{\dot 2}^1\\ \phi_{\dot 2}^2 \end{pmatrix}& -1 & 2 & 1 \\[4mm]
\delta_0 & 0 & 1 & 1 \\
\delta_1 & -2 & 1 & 1 \\
\eta & -\frac 23& 1 & 3 
\end{array}
$$
\caption{Scalar particle content with SM-representations}
\label{table:part-cont}
\end{table}
In terms of the original scalar fields $\phi_{\dot a}^b$ and $\Delta_{\dot aI}$ the vevs are of the following form:
\begin{align*}
\begin{pmatrix} \phi_{\dot a}^b \end{pmatrix} &= \begin{pmatrix} v & 0 \\ 0 & \sqrt{\mu^2/2 \Lambda_\Sigma}\end{pmatrix}\\
\begin{pmatrix} \Delta_{\dot a I} \end{pmatrix} &= \begin{pmatrix} w & 0 & 0 & 0 \\ 0 & 0 & 0 & 0 \end{pmatrix} .
\end{align*}
This shows that there are two scales of spontaneous symmetry breaking: at $10^{11}-10^{12} \GeV$ we have 
\begin{equation*}
SU(2)_R \times SU(2)_L \times SU(4) \to U(1)_Y \times SU(2)_L \times SU(3)
\end{equation*}
and then at electroweak scale (both $v$ and $\mu$) we have 
\begin{equation*}
U(1)_Y \times SU(2)_L \times SU(3) \to U(1)_Q \times SU(3)
\end{equation*}

\newcommand{\noopsort}[1]{}\def\cprime{$'$}

\end{document}